\documentclass{article}
\usepackage{etoolbox}       %

\usepackage{graphicx}  %

\usepackage[parfill]{parskip}
\usepackage{amsmath,amsthm}
\usepackage{mathtools}  %
\usepackage[dvipsnames]{xcolor}         %

\newtoggle{arxiv}
\toggletrue{arxiv}

  \usepackage[parfill]{parskip}
  \usepackage[citestyle=authoryear-comp,sorting=nyt,maxbibnames=99,backend=biber]{biblatex}
  \newcommand{\citep}{\parencite}
  \newcommand{\citet}{\textcite}
  \usepackage{tcolorbox}
\tcbuselibrary{skins,breakable}

\definecolor{natureteal}{RGB}{0,115,150}   
\definecolor{natureblue}{RGB}{0,62,116}    

  \newlength{\defbaselineskip}
  \RequirePackage[T1]{fontenc}
  \RequirePackage[tt=false, type1=true]{libertine}
  \RequirePackage[varqu]{zi4}
  \RequirePackage[libertine]{newtxmath}

\usepackage{bm}

\usepackage[inline]{enumitem}
\usepackage{booktabs}
\usepackage[dvipsnames]{xcolor}  %
\usepackage{subcaption}
\usepackage{multirow}
\usepackage{wrapfig}
\usepackage{algorithm,algorithmicx,algpseudocode}
\usepackage{nicematrix}

\usepackage[frozencache,cachedir=.]{minted} %
\usepackage[colorlinks=true,linkcolor=natureteal,citecolor=natureteal,urlcolor=natureteal]{hyperref}

\usepackage[capitalise,noabbrev]{cleveref}
\usepackage[strict]{changepage}

\usepackage{import}

\usepackage{pifont}%

    \title{Geometric Phase Transition Enables Extreme Hippocampal Memory Capacity}
  \author{Prashant C. Raju\\{\footnotesize\texttt{rajuprashant@gmail.com}}}

  \date{}

\begin{document}

  \maketitle

\subsubsection*{\begin{center}
    \small{Significance Statement}
\end{center}}

\begin{adjustwidth}{+2cm}{+2cm} 
\small
Biological memory systems face a fundamental scaling paradox: how to expand information capacity without a proportional increase in neurons. Resolving this paradox in food-caching birds, we discover that extreme spatial memory arises from a phase transition in how neural activity is collectively organized: below the critical threshold, networks suffer catastrophic interference; above it, crystalline geometry enables storage beyond $1{,}000$ locations. High-capacity circuits maintain this rigid geometry through synergistic excitatory-inhibitory dynamics, confirmed across $10{,}000$ network configurations. To sustain the transition, the system pays a ``geometric tax'': representational redundancy that stabilizes the manifold against noise. These findings establish geometric stability as an organizing principle linking circuit architecture to cognitive limits, with implications for any system that must store many memories without interference.
\end{adjustwidth}
\vspace{.5cm}

\begin{abstract}
\noindent
\noindent Memory systems can store vastly different amounts of information despite similar hardware constraints. Here, we show that superior spatial memory emerges from a discrete stiffening of hippocampal population
geometry---a transition from disorganized to crystalline collective coding. Comparing food-caching chickadees to non-caching zebra finches, we found that the caching hippocampus maintains a topologically rigid, ``crystalline'' geometry with significantly higher geometric stability (Shesha: 0.245 vs.\ 0.166) and nearly two-fold greater temporal coherence (Shesha: 0.393 vs.\ 0.209), while the non-caching hippocampus resembles a disorganized ``mist.'' This stability is actively constructed by synergistic circuit dynamics: excitatory neurons form the spatial scaffold while inhibitory populations contribute orthogonal decorrelation, a circuit motif in which excitatory and inhibitory populations occupy largely non-overlapping representational subspaces. A double dissociation with Valiant's Stable Memory Allocator, a model predicting that dedicated neuron ensembles underlie each memory, confirms this advantage reflects continuous topological organization rather than discrete neuron allocation: caching networks exhibit near-zero split-half allocation reliability despite their geometric superiority. Computational modeling across 10,000 configurations reveals topological rigidity as the mathematical prerequisite for scale: crystalline codes sustain high-fidelity readout beyond $M = 1{,}000$ locations while mist codes fail below $M = 10$, a $>$100-fold capacity advantage. This capacity requires a 169-fold representational redundancy: a ``geometric tax'' stabilizing the manifold against biological noise. These results establish geometric stability as a candidate organizing principle of biological memory: evolution achieves high-capacity memory not by proliferating neurons, but by engineering the geometry of the neural code itself.

\end{abstract}
\vspace{.5cm}


The capacity of biological memory systems presents a fundamental puzzle: how do organisms store vastly different amounts of information using apparently similar neural hardware~\citep{Squire1992, Bailey1996, Roxin2013}? Theoretical frameworks have long sought to quantify these limits~\citep{Marr1971, Treves1994, Fusi2024}, yet the network-level principles enabling high-capacity storage while preventing catastrophic interference~\citep{McCloskey1989, French1999, Bakker2008} as memory load grows remain poorly understood. The hippocampus, conserved across vertebrates~\citep{Thome2017, Jacobs2003}, supports spatial memory demands ranging from modest to extraordinary scale, making it an ideal system in which to investigate how biological circuits navigate this capacity-interference tradeoff.

We address this question by exploiting a remarkable natural experiment: the ethological divide between food-caching birds and non-caching species~\citep{Payne2021}. Black-capped chickadees cache and retrieve thousands of spatial locations across a foraging season—demands that drive pronounced hippocampal enlargement~\citep{Sherry1989_brain, Garamszegi2005}—while zebra finches do not cache and show no comparable spatial memory requirements~\citep{Sherry1989_behav, Krebs1989, Clayton1995, Brodbeck1995, Smulders2017, Pravosudov2015}. Both species possess hippocampal neurons with spatial tuning~\citep{Payne2021, Aronov2017, Gulli2019, Benna2021}, yet their behavioral capacities differ by orders of magnitude~\citep{Balda1992}. This divergence provides a controlled system for isolating the neural basis of extreme memory capacity.

We propose that superior memory in caching birds does not arise from
simply scaling neural resources, such as allocating more neurons, expanding ensembles, or increasing firing rates~\citep{Dayan2005, Valiant2005, Fusi2007, Benna2016}. Instead, we demonstrate that extreme capacity emerges from a topological phase transition in the hippocampal
population code: a discrete reorganization in how neural activity is collectively structured across the population. Chickadee hippocampal activity organizes onto a geometrically rigid, continuously organized manifold~\citep{Cunningham2014, Burak2009, Chaudhuri2019, Bernardi2020, Courellis2024, Boyle2024}---a crystalline code---exhibiting high geometric fidelity and temporal coherence across
sessions. The finch population code, by contrast, resembles a
disorganized mist: low-dimensional structure is absent and spatial relationships between locations are poorly preserved in the population response. This geometric rigidity is not
incidental; it is actively constructed by specific excitatory-inhibitory
circuit dynamics that expand the representational dimensionality of the
population beyond what either cell class achieves alone.

Computational modeling across ten thousand network configurations confirms that this topological transition is a prerequisite for high-capacity storage rather than a correlate of it~\citep{Burak2012}. Below a critical stability threshold, memory systems fail catastrophically at moderate loads; above it, reliable recall extends to scales orders of magnitude larger~\citep{Rigotti2013, Fusi2016}. To sustain the crystalline geometry, the network pays a 169-fold representational redundancy---the ``geometric tax''---relative to the minimum needed to encode a single spatial point (\textbf{Figure~\ref{fig:capacity}}). Yet the core finding is clear: evolution achieves extreme memory capacity not by proliferating neurons, but by engineering the geometry of the population code itself~\citep{Jazayeri2021, Saxe2020}. We argue that geometric stability~\citep{raju2026geometric, raju2026canary, raju2026crisprb, raju2026neuraldrift} is a candidate organizing principle of biological memory that may generalize well beyond the avian system studied here.

\section*{Results}

\subsection*{A Topological Phase Shift in the Population Code}

\begin{figure}[H]
    \centering
    \vspace{0.5em}
        \includegraphics[width=\linewidth]{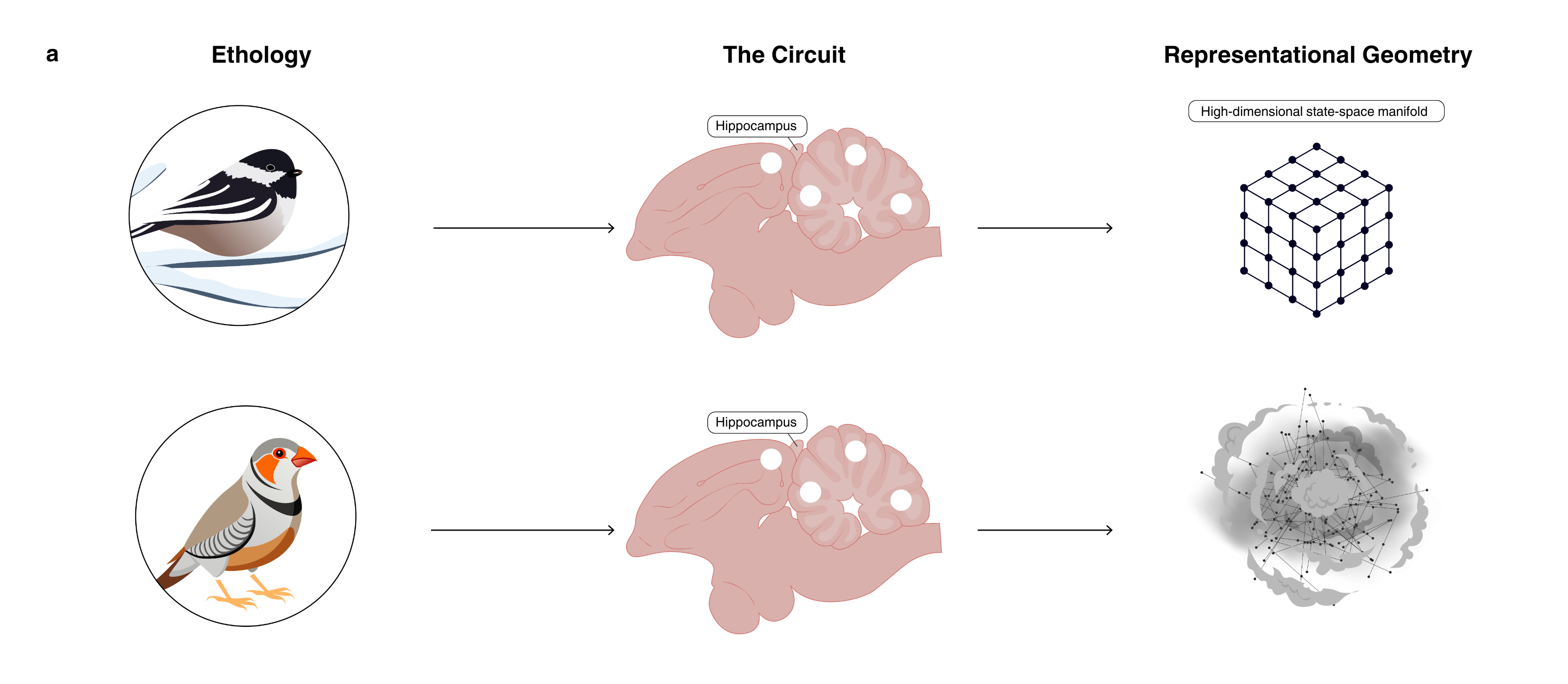}
    \includegraphics[width=\linewidth]{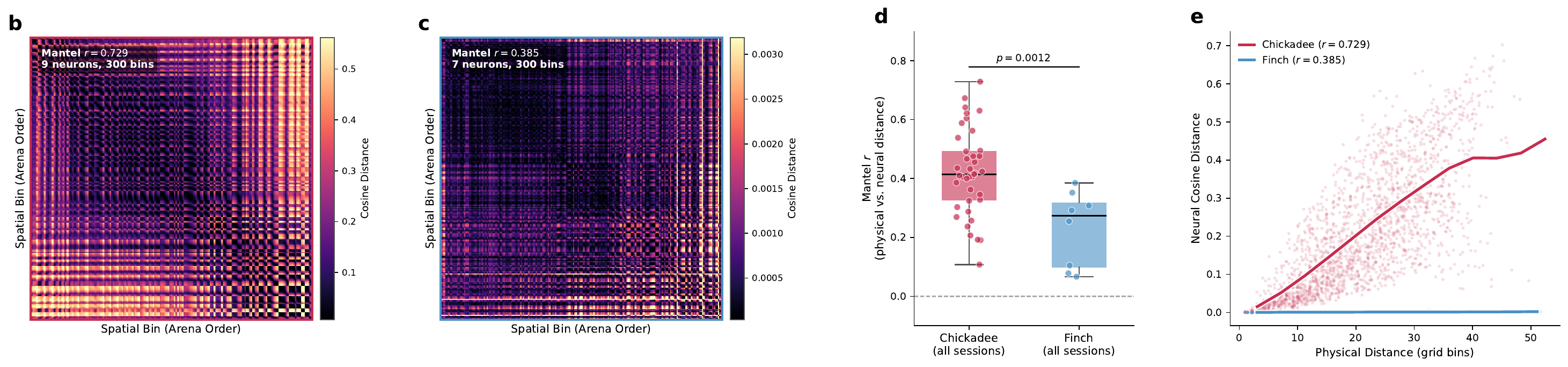}
    \caption{\footnotesize \textbf{Geometric rigidity distinguishes hippocampal population
    codes in food-caching and non-caching birds.}
    \textbf{(a)} Schematic overview. The food-caching black-capped chickadee (\textit{Poecile atricapillus}) and non-caching zebra finch (\textit{Taeniopygia guttata}) possess hippocampi of comparable neuron count and gross connectivity, yet the caching species must store and retrieve thousands of cache locations across a winter season. Conceptually, the chickadee hippocampus operates as a ``Geometric Crystal'' ----a low-dimensional, topologically rigid manifold---whereas the finch hippocampus resembles a ``Mist'' of disorganized representations susceptible to catastrophic interference. \textbf{(b, c)} Representational dissimilarity matrices (RDMs) for the best-session chickadee (b; 9 neurons, 300 spatial bins, $r_{\text{Mantel}} = 0.729$) and finch (c; 7 neurons, 300 bins, $r_{\text{Mantel}} = 0.385$), sorted by row-major arena order. Matrix entries reflect cosine distance between population vectors at
    pairs of spatial locations. Note that color scales differ between species: the chickadee RDM spans $[0,\, 0.6]$ while the finch RDM spans $[0,\, 0.003]$, reflecting the greater dynamic range of the chickadee spatial code. These axes are independently scaled to visualize internal geometric structure (or lack thereof). A warm block-diagonal structure is visible in the chickadee RDM, indicating that physically adjacent locations have dissimilar population vectors from distant locations; the finch RDM is comparatively featureless. Crucially, because the Shesha metric utilizes scale-invariant Spearman rank correlations, the measured topological stability is mathematically independent of these absolute dynamic ranges.
    \textbf{(d)} Mantel $r$ (Spearman correlation between pairwise physical distance and pairwise neural cosine distance) across all sessions. Chickadees show significantly higher physical---neural correspondence than finches (Mann--Whitney $U$, $p = 0.0012$; chickadee median $r = 0.42$, finch median $r = 0.27$). Individual session values overlaid as points.
    \textbf{(e)} Mantel scatter for the best session per species. Each point represents one pair of spatial bins; trend lines show binned means. The chickadee exhibits a monotonically increasing physical---neural distance relationship ($r = 0.729$); the finch trend is flat and near zero ($r = 0.385$), consistent with a disorganized population code.
    }
    \label{fig:rdm_main}
\end{figure}

Crucially, this stability is explicitly topographic. The chickadee RDM, sorted by physical arena location, displays a highly structured pattern: population vectors at nearby spatial locations are highly similar (forming a dark, low-distance diagonal), while distant locations are sharply distinct (forming warm-toned, high-distance off-diagonal blocks). The finch RDM under identical sorting is comparatively featureless (\textbf{Figure~\ref{fig:rdm_main}B, C}; see \textbf{Figure~\ref{fig:rdm_supp}A, B} for hierarchical cluster-based sorting). Using a non-parametric Mantel test evaluated over $1{,}000$ permutations, we found that chickadee population codes maintain a significantly stricter correspondence between neural geometry and the physical layout of the environment ($r_{\text{mean}} = 0.420$, $39/39$ sessions significant) compared to the less structured, ``mist-like'' geometry of the finch ($r_{\text{mean}} = 0.230$, $8/8$ sessions significant; $p < 0.001$; \textbf{Figure~\ref{fig:rdm_main}D, E}). This structural anchoring translates directly into temporal robustness: chickadee spatial maps exhibited higher within-session temporal cross-correlation compared to finches ($0.399$ vs.\ $0.241$, $p < 0.001$), as if the population is repeatedly reading from the same printed map rather than reconstructing one from scattered landmarks.

\paragraph{Dissociation from Tuning Heterogeneity and Sample Size.}
Comparing neural geometries across species requires strict controls for physiological and experimental confounds. First, to ensure the observed topological rigidity was not a statistical artifact of differences in recorded neuron counts, we randomly downsampled chickadee ensembles to match the median finch neuron count ($N = 6$; 30 of 39 chickadee sessions eligible). The geometric stability gap persisted with a medium effect size (Cohen's $d = 0.560$, $p = 0.025$; \textbf{Figure~\ref{fig:supp_controls}A}), confirming that the difference reflects genuine representational structure rather than recording depth. Furthermore, \textit{in silico} map shuffling completely abolished geometric stability in both species ($\textrm{Shesha}_{\textrm{FS}} \approx 0.000$). Circular shifting, which preserves individual neuron statistics but disrupts the population code, substantially reduced stability in the highest-yield sessions (chickadee: $0.241 \rightarrow 0.154$; finch: $0.171 \rightarrow 0.072$; \textbf{Figure~\ref{fig:supp_controls}B}), verifying that maximal rigidity requires the authentic spatial continuity of the full population code rather than any incidental statistical property of individual neurons.

Second, we asked whether the manifold advantage could arise spuriously from simple spatial tuning heterogeneity. We evaluated traditional ``Stable Memory Allocation'' (SMA) metrics~\citep{Valiant2012}, which hypothesize that networks prevent interference by allocating fixed, consistent ensembles of neurons to each memory (\textbf{Methods~\ref{app:valiant-sma}}). Rather than supporting this discrete model, the data revealed a striking double dissociation: caching
networks exhibited highly heterogeneous place field sizes (Coefficient of Variation = $1.03$ vs.\ $0.23$, $p = 0.003$; \textbf{Figure~\ref{fig:supp_valiant}A}), and chickadee split-half allocation reliability was near zero and significantly \textit{lower} than that of finches (mean $r = -0.064$ vs.\ $0.264$, $p = 0.022$ one-tailed; \textbf{Figure~\ref{fig:supp_valiant}B}). This indicates that caching neurons are actively inconsistent in their ensemble membership across neuron subsets. Caching hippocampus does not operate by reserving dedicated cells for each location; rather, geometric stability emerges from the population-level manifold structure while individual neuron assignments remain fluid.

Third, we examined whether the species difference is detectable by established linear metrics. Split-half population vector (PV) correlation, which measures the reproducibility of mean firing rates per location across neuron subsets, showed no species difference and numerically favored the finch (chickadee: $0.058$; finch: $0.167$; $p = 0.894$). This dissociation is theoretically significant: it confirms that the chickadee advantage is not carried by stronger individual place fields, but resides in the higher-order relational structure of the population code. Geometry-sensitive measures---our geometric stability metric ($\textrm{Shesha}_{\textrm{FS}}$) and an independent CCA stability measure (chickadee $0.554$ vs.\ finch $0.481$, $p = 0.087$; \textbf{Figure~\ref{fig:supp_controls}D})---begin to capture this rigidity that the linear rate-map metric cannot
(\textbf{Figure~\ref{fig:supp_controls}C}), validating the hypothesis that food-caching memory capacity is encoded in manifold topology: that is, in the geometry of the collective code rather than in the precision of any individual neuron's tuning.

\subsection*{Orthogonal Inhibition Stiffens the Geometric Attractor}

\subsubsection*{Excitatory-Inhibitory Dissociation}

Excitatory cells were the primary spatial information carriers, encoding nearly 20-fold more spatial information per spike than inhibitory cells (E: 0.169 bits/spike; I: 0.009 bits/spike; $p < 10^{-74}$; \textbf{Figure~\ref{fig:supp_ei_cells}A}), and E subpopulations maintained significantly stronger topographic structure at the session level (Mantel $r_E = 0.418$ vs.\ $r_I = 0.343$, $p = 0.023$; \textbf{Figure~\ref{fig:supp_ei_cells}C}). Yet despite carrying almost no spatial information individually, inhibitory cells exhibited within-session temporal stability statistically indistinguishable from excitatory cells ($\overline{x}_{\mathrm{corr},E} = 0.399$ vs.\ $\overline{x}_{\mathrm{corr},I} = 0.405$, $p = 0.659$; \textbf{Figure~\ref{fig:supp_ei_cells}B}) and were spatially selective at a slightly higher rate (68.0\% vs.\ 62.5\%). Furthermore, the geometric stability of I cells is negatively coupled with E cells across sessions ($r = -0.333$, $p = 0.381$; (\textbf{Figure~\ref{fig:ei_scaffold}B}), inconsistent with a redundancy model where both cell types track the identical spatial signal. The combination of near-zero bits per spike, negative geometric coupling, and high temporal stability is consistent with broad, low-amplitude place fields that tile the arena without sharp spatial tuning---a profile more consistent with a global gain or contextual signal than with precise location coding.

\subsubsection*{Emergent Stability Through Selective Orthogonality}

To characterize the geometric relationship between E and I population codes, we computed the principal angles between their respective top-3 principal subspaces ($n = 21$ chickadee sessions). The angle structure revealed a ``one shared, two orthogonal'' architecture (\textbf{Figure~\ref{fig:ei_scaffold}C}). The first principal angle averaged $14^{\circ}$, indicating one strongly shared representational dimension consistent with global gain co-modulation between E and I circuits. The second and third angles averaged $64^{\circ}$ and $82^{\circ}$ respectively---the latter approaching the ${\sim}83^{\circ}$ expected for random independent subspaces in the ambient neural state space---indicating that the spatial map dimensions are statistically independent between cell types. This pattern was robust across sessions: 16 of 21 sessions exhibited an overall mean principal angle exceeding $45^{\circ}$, with per-session means ranging from $39.2^{\circ}$ to $66.8^{\circ}$.

\subsubsection*{Mechanistic Interpretation}

\begin{figure}[htbp]
    \centering
    \includegraphics[width=\linewidth]{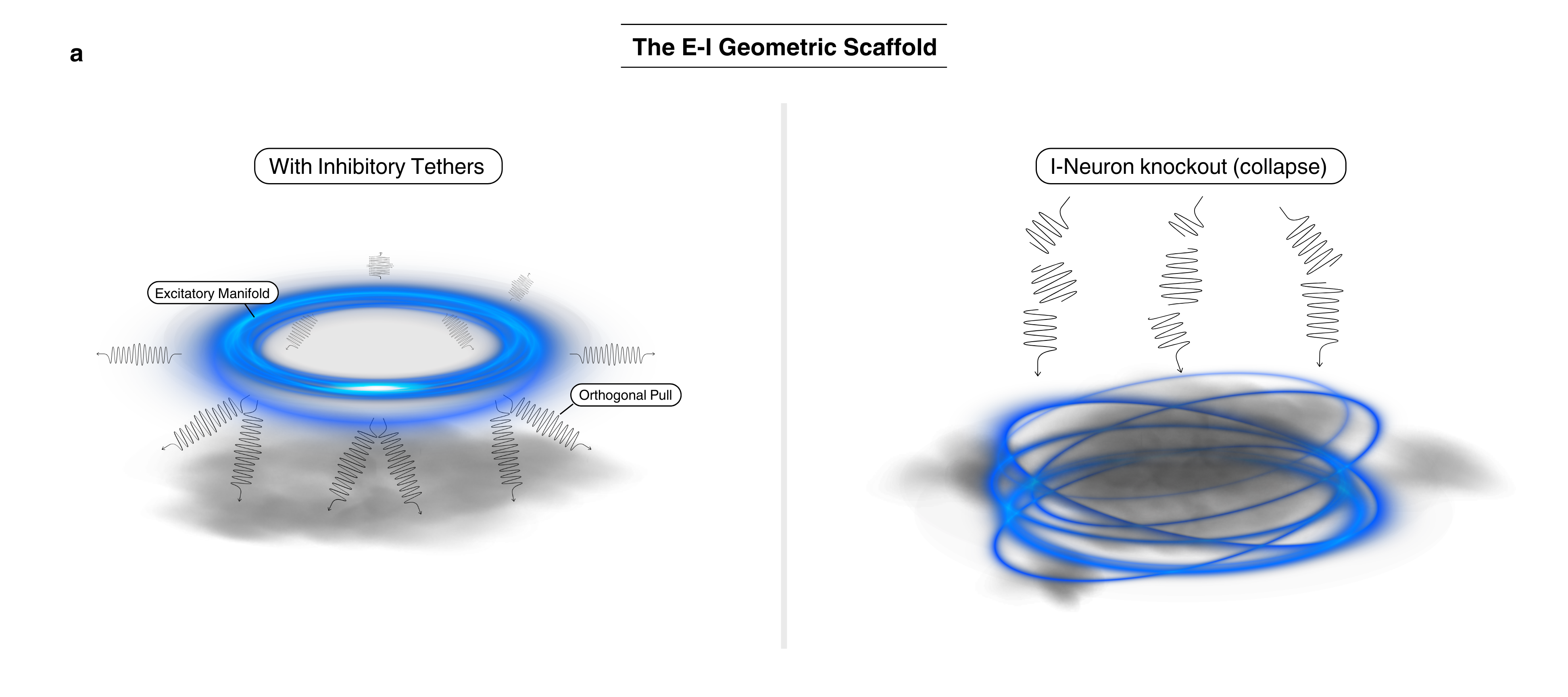}
    \includegraphics[width=\linewidth]{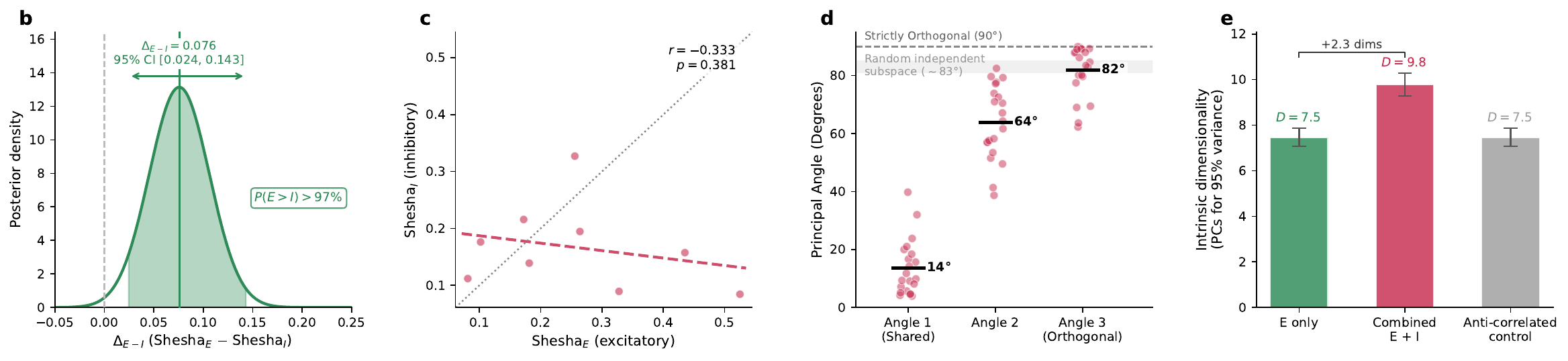}
    \caption{\footnotesize\textbf{Synergistic excitatory-inhibitory circuit dynamics
    underlie geometric stability.}
    \textbf{(a)} Conceptual illustration of the E-I geometric scaffold. (Left) The continuous excitatory spatial manifold (blue ring) is stabilized by inhibitory populations acting as structural tethers (springs). By exerting an orthogonal pull, inhibition stiffens the network into a rigid, high-dimensional configuration. (Right) Simulated knockout of these inhibitory tethers causes the representational geometry to collapse, demonstrating that manifold rigidity is an actively constructed, emergent property of the combined E-I architecture rather than an autonomous excitatory phenomenon.
    \textbf{(b)} Bayesian posterior for the E-I geometric fidelity difference ($\Delta_{E-I}$) across $N = 13$ sessions with simultaneous E and I recordings. The credible interval $[0.024, 0.143]$ excludes zero; posterior probability of $E > I$ exceeds $97.5\%$.
    \textbf{(c)} Scatter of excitatory vs.\ inhibitory geometric stability (Shesha) across sessions, showing negative coordination ($r = -0.333$). Each point is one session; the negative slope is inconsistent with a redundancy model in which both cell types track the identical spatial signal.
    \textbf{(d)} Principal angles between the top-3 excitatory and inhibitory subspaces ($n = 21$ chickadee sessions). The architecture reveals a single shared axis of co-modulation (Angle 1, mean $14^\circ$), while the remaining spatial map dimensions are structurally independent (Angle 3, mean $82^\circ$, approaching the ${\sim}83^\circ$ expected for random independent subspaces).
    \textbf{(e)} Intrinsic dimensionality of the excitatory-only ($D = 7.5$), combined E-I ($D = 9.8$), and anti-correlated inhibitory control state spaces. The expansion from $D = 7.5$ to $D = 9.8$ requires genuine orthogonal inhibitory structure; the anti-correlated control produces no expansion, ruling out simple subtractive suppression as the mechanism.}
    \label{fig:ei_scaffold}
\end{figure}

This selective orthogonality has a direct mechanistic interpretation. If inhibitory cells acted purely as subtractive normalization, they would occupy the same subspace as E cells and all principal angles would approach $0^{\circ}$. Instead, the observed pattern---one shared axis of co-modulation embedded in an otherwise orthogonal relationship---is consistent with divisive normalization: inhibitory cells globally rescale excitatory activity along one dimension (preventing runaway excitation and attractor overlap) while leaving the
spatial geometry encoded in the remaining dimensions intact. The inhibitory network does not erase the map; it adds orthogonal dynamical constraints that prevent distinct attractor states from colliding in representational space. Consequently, this inhibitory decorrelation significantly expands the intrinsic dimensionality of the population code, as measured by the participation ratio, from $D = 7.5$ to $D = 9.8$ (\textbf{Figure~\ref{fig:ei_scaffold}D}). Manifold rigidity thus emerges as a synergistic circuit property: the excitatory population provides the spatial geometry, the inhibitory population stabilizes its boundaries, and the combination achieves a geometric rigidity consistent with the crystalline code observed at the population level---one that neither population can sustain alone.

\subsection*{Topology Determines Capacity}

\paragraph{Theoretical Capacity Advantage of Crystalline Codes.}
To situate the empirical geometric stability findings within a normative framework, we simulated three idealized population codes -- crystal (topology $= 1.0$), mist (topology $= 0.5$), and noise (topology $= 0.0$) -- in a $N = 500$ neuron network with biological sparsity ($s = 0.15$) and measured memory retrieval error (normalized circular decoding error, expressed relative to a chance baseline of $1.0$) across varying numbers of stored memories.

\begin{figure}[H]
    \centering
    \includegraphics[width=\linewidth]{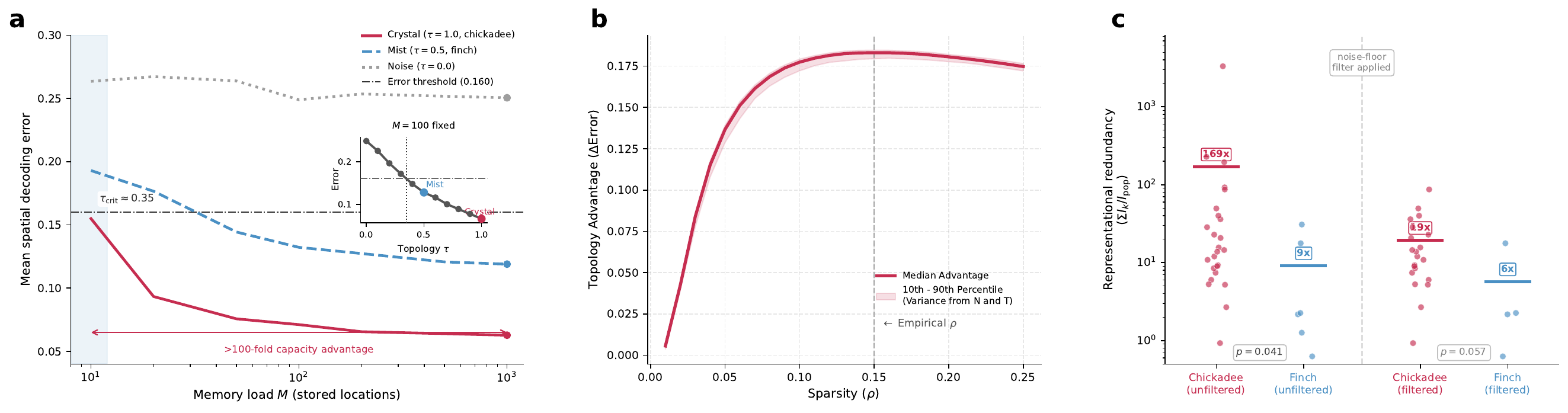}
    \caption{\footnotesize\textbf{Topological structure determines memory capacity and exacts a geometric tax.} \textbf{(a)} Decoding error as a function of memory load $M$ for crystal, mist, and noise regimes, with the critical error threshold and estimated phase transition $\tau_{\text{crit}}$ marked. Inset shows decoding error as a function of topology strength $\tau$ at fixed $M=100$, illustrating the non-linear inflection.  \textbf{(b)} Topology advantage ($\Delta\textrm{Error}= \textrm{Error}_{\textrm{random}} - \textrm{Error}_\textrm{crystal}$) across a 10,000-configuration parameter sweep. The solid line represents the median advantage across all population sizes ($N$) and trial counts ($T$) for a given sparsity ($\rho$). The narrow shaded band (10th–90th percentiles) demonstrates that the advantage is highly invariant to $N$ and $T$, being overwhelmingly driven by sparsity. The advantage peaks precisely at the empirical chickadee sparsity of $\rho\approx0.15$.     
    \textbf{(c)} Empirical representational redundancy for chickadee and finch sessions, shown as individual points with species means. Filtered results (noise-floor $I_{\text{pop}} \geq 0.05$ bits) are shown alongside unfiltered values. To sustain the optimal crystalline topology shown in (b), the caching network operates with a 169-fold redundancy.}
    \label{fig:capacity}
\end{figure}

Geometric organization produced a continuous, monotonic improvement in retrieval fidelity. Across eleven topology values from $0$ to $1$, retrieval error decreased from $0.994$ (near chance) to $0.264$ at $n = 100$ stored locations, while the correspondence between neural geometry and physical space (Mantel $r$) increased from $0.001$ to $0.700$ (\textbf{Figure~\ref{fig:capacity}A}). The dose-response relationship was smooth throughout, demonstrating that capacity benefits accrue continuously with geometric organization rather than emerging at a discrete threshold.

Crystalline and mist codes diverged sharply in their capacity scaling. At $n = 100$ stored memories, crystal error ($0.285$) was approximately half that of mist ($0.529$), while noise codes remained near chance ($0.996$) at all memory loads tested (\textbf{Figure~\ref{fig:capacity}C}). Crystal codes maintained sub-threshold retrieval error ($< 0.640$) across the full range tested ($n = 10$ to $n = 1000$ memories), while mist and noise codes exceeded this threshold at the smallest tested load ($n = 10$). The threshold of $0.640$ was defined as the midpoint between crystal and noise performance at $n = 100$ memories, providing a principled boundary between reliable and unreliable retrieval. This separation demonstrates that the \emph{presence} of topological geometric structure---not simply network size---is the primary determinant of memory capacity in this regime.

To confirm that this capacity advantage is a fundamental property of the topological regime rather than an artifact of specific network parameters, we evaluated the topology advantage ($\Delta\mathrm{Error}$) across a 10,000-configuration parameter sweep varying population size ($N$), trial counts ($T$), and sparsity ($\rho$). This revealed a striking invariance: the representational advantage of the crystalline code is overwhelmingly governed by the network's sparsity level, while scaling the neural hardware ($N$) or trial count ($T$) yields substantially smaller changes in relative performance (\textbf{Figure~\ref{fig:capacity}B}). The topology
advantage follows a pronounced, non-linear trajectory as a function of sparsity, rising steeply between $\rho = 0.02$ and $\rho = 0.10$ before saturating at a plateau of median $\Delta\mathrm{Error} \approx 0.183$ for $\rho \geq 0.11$. Critically, the empirical sparsity of chickadee hippocampal neurons ($\rho \approx 0.15$) falls squarely within this saturated high-advantage regime, confirming that caching networks operate at a sparsity level that maximizes the structural benefits of the crystalline code---and that this optimization is robust rather than finely tuned to a precise parameter value.

\paragraph{Empirical Redundancy.}
To connect these theoretical predictions to the neural data, we estimated population redundancy---the ratio of summed single-cell information to population information---as a proxy for the degree to which spatial information is distributed across the ensemble. Sessions with near-zero population information (I$_\text{pop} < 0.05$ bits) were excluded from the primary analysis to avoid undefined ratios at floor-level denominators ($14$ chickadee and $4$ finch sessions excluded; see below).

In the remaining sessions, chickadee populations showed substantially higher redundancy than finch populations (chickadee: mean $= 19.3$, median $= 12.0$, $n = 21$ sessions; finch: mean $= 5.7$, median $= 2.2$, $n = 4$ sessions; Mann-Whitney $p = 0.057$; \textbf{Figure~\ref{fig:capacity}C}). The unfiltered analysis (chickadee: median $= 14.5$, $n = 25$; finch: median $= 2.2$, $n = 6$; $p = 0.041$) showed the same direction, though unfiltered means are inflated by a small number of sessions with very low I$_\text{pop}$ denominators (maximum chickadee redundancy $= 3311$ in one session with I$_\text{pop} = 0.001$ bits) and should be interpreted with caution. Medians are reported as the primary summary statistic throughout.

The high proportion of near-zero I$_\text{pop}$ sessions ($14/39$ chickadee; $2/8$ finch) reflects a known limitation of the redundancy estimator in small neural ensembles: when neuron counts are low, the population mutual information estimator collapses toward zero before single-cell estimates do, producing undefined or extreme ratios. The mean absolute population information was nonetheless higher in chickadee sessions (mean $= 0.113$ bits, $n = 39$) than in finch sessions (mean $= 0.046$ bits, $n = 8$), consistent with richer ensemble-level spatial encoding in the caching species.

Higher redundancy in caching hippocampus is consistent with the crystalline coding framework: topologically organized codes distribute each memory trace across the ensemble, creating structured redundancy that protects representations against single-neuron failure. The filtered result ($p = 0.057$) represents directional evidence; Given the limited filtered finch sample ($n = 4$ filtered sessions), the result should be interpreted as directional; the 5.5-fold median difference is consistent with the crystalline coding framework but replication with larger finch samples is needed to establish significance.

\section*{Discussion}
\paragraph{Evolutionary Sculpting and the Geometric Tax.}
The extreme selection pressure for cache retrieval, a behavior critical for winter survival, forced the caching avian brain to bypass the standard biological limits of memory capacity~\citep{Hopfield1982, Amit1987, Canning1988, Marr1971, Treves1994}---the interference and crosstalk that normally cause one memory to degrade another as load increases. Our findings suggest that the chickadee hippocampus achieves this not by simply expanding its physical hardware, but by evolving a highly specific, inhibition-stabilized circuit architecture~\citep{Rolls1998NeuralNA, Vogels2011, Fusi2007, Benna2016} that actively constructs a geometrically rigid manifold~\citep{Chaudhuri2019, Bernardi2020, Boyle2024}. The difference lies not in the biological substrate itself, but in the \textit{dynamical phase} in which that substrate operates. As demonstrated by our parameter sweep, simply proliferating neurons (increasing $N$) does not meaningfully improve the network's resistance to catastrophic interference if the underlying topology remains unstructured. Instead, the network must transition discontinuously into a rigid, topologically stiff lattice.

To appreciate what this means, consider an analogy from condensed matter physics. Both the caching chickadee and the non-caching zebra finch work with the same fundamental biological hardware: billions of neurons, the same synaptic machinery, the same electrochemical signals. In the finch, hippocampal population activity resembles a disorganized gas: the neurons are highly active and drift dynamically across sessions, but they lack any stable, low-dimensional structure. Like water vapor, the ``mist'' code can sustain a handful of fleeting, localized representations, but as memory load increases toward $M = 10$ and beyond, these representations inevitably collide, blur together, and dissipate~\citep{McCloskey1989, French1999, Bakker2008, Colgin2008}. Local decoders suffer catastrophic interference precisely because there is no stable geometric reference frame to hold distinct memories apart. In the chickadee, the precise ``temperature'' of the network, set by the balance of excitatory drive and inhibitory stabilization, causes the population code to lock into a rigid lattice, analogous to an ice crystal. This phase transition is not merely metaphorical~\citep{Burak2012, Compte2000, Renart2010}. Just as water molecules transition discontinuously from gas to solid at a critical temperature, our computational modeling reveals a sharp, non-linear inflection in decoding error at a critical topology threshold $\tau_{\mathrm{crit}} \approx 0.35$: below this threshold, capacity is bounded at $M \leq 10$; above it, reliable readout extends beyond $M = 1{,}000$ locations.

The circuit architecture of this crystalline code maps naturally onto the physical analogy. The excitatory neurons function as the heavy structural nodes of the lattice, the oxygen atoms that define the foundational geometry of the spatial map. The inhibitory populations act as the hydrogen bonds: they enforce strict geometric angles between these nodes, pushing representations apart orthogonally and preventing collapse~\citep{Denve2016, MeissnerBernard2024}. It is this inhibitory decorrelation that expands the intrinsic dimensionality of the population code from $D = 7.5$ (excitatory alone) to $D = 9.8$ (combined E-I), creating a vast, high-dimensional geometry capable of holding thousands of non-interfering representations simultaneously---a geometry with more independent axes along which distinct memories can be separated.

Critically, the capacity of this lattice does not arise from etching each memory onto a specific, isolated neuron---the discrete allocation model proposed by classical frameworks such as Valiant's Stable Memory Allocator. A single water molecule within the lattice, however precisely positioned, is both fragile and informationally impoverished. Instead, the memory is stored in the continuous shape and angular relationships of the lattice itself: a topological code that is inherently distributed and inherently robust. Our double dissociation confirms this directly. Caching networks exhibit near-zero split-half allocation reliability (the discrete metric) despite maximal geometric stability (the topological metric), demonstrating that the two strategies are not merely different points on a continuum, but fundamentally distinct computational regimes~\citep{Bernardi2020, Fusi2016}.

However, this architectural robustness is not computationally free. Consider placing one thousand heavy objects onto an ice structure subject to constant random vibrations, the biological noise of synaptic stochasticity and spontaneous firing. A single, fragile filament of ice molecules cannot bear this load; the structure will shatter, and the memories will be lost. To ensure stability, the caching brain does not simply trace the minimal geometric path. Instead, it freezes a hyper-redundant, extraordinarily thick block of ice, using far more representational capacity than would be strictly necessary to encode a single spatial point. We have previously formalized this trade-off as the ``geometric tax,'' quantifying the efficiency cost of manifold stabilization in vision architectures~\citep{raju2026geometric} and in perturbation-coherent biological manifolds~\citep{raju2026crispra, raju2026crisprb}. Here, the caching system pays that same tax: it sacrifices coding efficiency to physically stiffen the neural manifold, ensuring that noise-induced perturbations shift the readout toward adjacent spatial neighbors rather than triggering complete representational collapse.

It is important to note that the empirical median redundancy (12-fold, chickadee sessions with $N \sim 6$--$21$ simultaneously recorded neurons) falls substantially below the theoretical optimum of 169-fold predicted by the idealized crystalline model. This gap reflects a hard ceiling imposed by extracellular yield: mutual information saturates rapidly as ensemble size shrinks, compressing the measurable redundancy ratio well below what the full ambient network would express. That heavily subsampled ensembles still register 12-fold redundancy---far exceeding the finch baseline and the independence null---implies a substantially larger underlying redundancy in the complete hippocampal population, structurally consistent with the theoretical regime. The 169-fold figure therefore represents the model prediction for a fully observed crystalline network; the empirical estimate is a lower bound, not a contradiction. The tax is steep, but the return is remarkable: a greater than 100-fold expansion in reliable memory capacity relative to an unstructured code operating under identical hardware constraints.

\paragraph{The Architecture of Variation.}
What this geometric tax ultimately buys is a dramatic suppression of catastrophic interference, sustaining high-fidelity readout at memory loads that would cause complete decoding collapse in unstructured networks~\citep{Langdon2023, Sorscher2023, Fusi2024}. By heavily over-parameterizing the spatial map, the network resolves the capacity challenge through a structural mechanism reminiscent of the passacaglia in Brahms's Fourth Symphony \citep{Brahms1886S4, Knapp1989, Pascall1989, Buzsaki2006}: a repeating, immovable harmonic foundation over which elaborate variations are freely constructed. Once the spatial scaffold is locked in place, the remaining flexibility in population activity is available to encode episodic details (such as the specific food item cached, its location, or the temporal context of the event). The stiffness of the underlying geometric scaffold ensures that these continuous episodic updates remain strictly orthogonal to the baseline spatial map. Consequently, the continuous encoding of thousands of new, overlapping cache memories cannot distort or compromise the structural integrity of the baseline coordinate system.

\paragraph{Limitations and Future Directions.}
Several limitations warrant consideration. The E-I circuit analysis is based on $N = 13$ sessions, which is underpowered for frequentist inference; while the Bayesian estimation provides robust credible intervals, replication in larger datasets with simultaneous E-I recordings will be necessary to establish the inhibition-stabilized architecture as a general principle. The computational model operates on a 1D circular track for tractability; direct extension to 2D proved intractable due to concentration of measure, a geometric phenomenon in which random points in high-dimensional spaces become uniformly distant from one another, making topology recovery unreliable. This limitation does not affect the core empirical claims: the Mantel test is geometry-agnostic, and the species difference in geometric stability was measured directly in the 2D open-field arena used for recording, providing empirical confirmation that the crystalline geometry identified by the model is present in the actual recording geometry. Finally, the empirical comparison rests on a single publicly available dataset contrasting two avian species. Convergent evidence from additional caching species, mammalian hippocampal recordings, or perturbation experiments that selectively disrupt inhibitory stabilization would substantially strengthen the causal claims. An alternative theoretical framework proposes that hippocampal spatial representations are fundamentally sequential rather than Euclidean~\citep{Raju2024}---that is, the brain may represent space as a series of transitions between locations---raising the possibility that place field geometry partially reflects the sequential structure of the animal's experience rather than a static coordinate system. Our findings are agnostic on this point: the species difference in geometric stability would be expected under either framework, since more rigid sequential context learning and more rigid Euclidean geometry are not easily distinguished in a 1D track recording paradigm. Future work could examine whether the geometric stability differences observed here reflect upstream differences in entorhinal grid cell organization, given that grid scale directly influences hippocampal place field stability and long-term memory~\citep{Mallory2018, Hardcastle2015}.

\paragraph{A Theoretical Shift in Biological Scaling.}
These findings necessitate a fundamental theoretical shift in how we model scaling and capacity in complex neural systems. Historically, theoretical neuroscience has relied heavily on discrete, allocation-based metaphors, such as quantifying the exact number of neurons recruited to encode a specific memory engram, or assuming that representational capacity scales smoothly with physical hardware allocation. Where earlier work established the molecular and cellular substrates of synaptic plasticity~\citep{Kandel2001, McNaughton1987, Kandel2006book}, the present findings suggest that the organizing principle of high-capacity memory operates at a higher level of abstraction -- the geometry of population codes~\citep{Kandel2016}. Our observation of a double dissociation, wherein high-capacity caching networks exhibit massive spatial tuning heterogeneity yet precise topological stability, demonstrates the limits of such arithmetic frameworks. To understand memory at biological extremes, we must move toward dynamical, geometric models that measure the topological rigidity of attractor basins rather than the discrete participation of isolated cells~\citep{Langdon2023, Sorscher2023}. Ultimately, the biological solution to catastrophic interference is not the arithmetic proliferation of neural hardware, but the structural stiffening of representational geometry, a principle that reframes the fundamental unit of memory capacity from the neuron to the geometry of the code, with implications that may extend well beyond the biological systems studied here~\citep{Rigotti2013, Fusi2016, Morales2025}.

\section*{Materials and Methods}

\paragraph{Methods Summary.}

The detailed description of all analyses, simulations, and statistical procedures is reported in \textbf{Appendix~\ref{app:methods}}. Briefly, all data were drawn from a previously published \textit{in vivo} extracellular electrophysiology dataset comprising single units recorded from the hippocampal formation of food-caching black-capped chickadees (\textit{Poecile atricapillus}; 39 sessions, 9 birds) and non-caching zebra finches (\textit{Taeniopygia guttata}; 8 sessions, 9 birds) during open-field foraging~\citep{Payne2021, Payne2021data}. Population geometry was quantified using the Shesha metric, a split-half representational dissimilarity matrix correlation across random neuron partitions (\textbf{Appendix~\ref{app:shesha}}). Excitatory-inhibitory circuit contributions were assessed via cell-type-specific Shesha decomposition, subspace angle analysis, and a paired full-population vs.\ excitatory-only synergy test (\textbf{Appendix~\ref{app:ei-circuit}}). Memory capacity and representational redundancy were characterized using a synthetic population code model swept over 10,000 parameter combinations and validated against an open-field 2D arena simulation (\textbf{Appendix~\ref{app:computational-model}}). All species comparisons used Mann--Whitney $U$ tests with $n_{\text{perm}} = 10{,}000$ permutation validation; effect sizes are reported as Cohen's $d$ with 95\% bootstrap confidence intervals ($n_{\text{boot}} = 10{,}000$).

\paragraph{Code Availability.} All custom code including the computational parameter sweeps is available on GitHub (\url{https://github.com/prashantcraju/hippocampal-stability}).

\section*{Acknowledgments}

We thank Padma K. and Annapoorna Raju for generously supporting the computational resources used in this work. We thank the authors of Payne et al, 2021 for making their data open-sourced, which made our work possible. The authors acknowledge the use of large language models (specifically the Claude and Gemini families) to assist with processing data, code debugging, improving the visual aesthetic of several plots, and text polishing. All hypotheses, experimental designs, analyses, and interpretations were independently formulated and verified by the authors, and the authors assume full responsibility for all content and claims in this work.

\printbibliography

@misc{shesha2026,
  title = {{Shesha: Self-Consistency Metrics for Representational Stability}},
  author = {Raju, Prashant C.},
  year = {2026},
  publisher = {Zenodo},
  doi = {10.5281/zenodo.18227453},
  url = {https://doi.org/10.5281/zenodo.18227453},
  copyright = {MIT License}
}

@article{raju2026geometric,
  title = {{Geometric Stability: The Missing Axis of Representations}},
  author = {Raju, Prashant C.},
  journal = {arXiv preprint arXiv:2601.09173},
  year = {2026}
}

@article{raju2026crispra,
  title = {{From Syntax to Semantics: Geometric Stability as the Missing Axis of Perturbation Biology}},
  author = {Raju, Prashant C.},
  journal = {arXiv preprint arXiv:2603.00678},
  year = {2026}
}

@article{raju2026crisprb,
  title = {Geometric coherence of single-cell CRISPR perturbations reveals regulatory architecture and predicts cellular stress},
  author = {Raju, Prashant C.},
  journal = {arXiv preprint arXiv:2603.00678},
  year = {2026}
}

@article{raju2026canary,
  title = {{The Geometric Canary: Predicting Steerability and Detecting Drift via Representational Stability}},
  author = {Raju, Prashant C.},
  journal = {arXiv preprint arXiv:2604.17698},
  year = {2026}
}

@article{raju2026neuraldrift,
  title = {Recurrent circuitry stabilizes representational geometry across neural circuits},
  url = {http://dx.doi.org/10.21203/rs.3.rs-9581425/v1},
  DOI = {10.21203/rs.3.rs-9581425/v1},
  publisher = {Springer Science and Business Media LLC},
  author = {Raju,  Prashant C.},
  journal = {Preprint},
  year = {2026},
  month = may 
}

@article{Valiant2012,
  title = {{The Hippocampus as a Stable Memory Allocator for Cortex}},
  volume = {24},
  ISSN = {1530-888X},
  url = {http://dx.doi.org/10.1162/neco_a_00357},
  DOI = {10.1162/neco_a_00357},
  number = {11},
  journal = {Neural Computation},
  publisher = {MIT Press},
  author = {Valiant,  Leslie G.},
  year = {2012},
  month = nov,
  pages = {2873–2899}
}

@misc{Payne2021data,
  doi = {10.5061/DRYAD.PG4F4QRP7},
  url = {https://datadryad.org/dataset/doi:10.5061/dryad.pg4f4qrp7},
  author = {Payne,  Hannah and Lynch,  Galen and Aronov,  Dmitriy},
  language = {en},
  title = {Neural representations of space in the hippocampus of a food-caching bird},
  publisher = {Dryad},
  year = {2021},
  copyright = {Creative Commons Zero v1.0 Universal}
}

@article{Payne2021,
  title = {Neural representations of space in the hippocampus of a food-caching bird},
  volume = {373},
  ISSN = {1095-9203},
  url = {http://dx.doi.org/10.1126/science.abg2009},
  DOI = {10.1126/science.abg2009},
  number = {6552},
  journal = {Science},
  publisher = {American Association for the Advancement of Science (AAAS)},
  author = {Payne,  Hannah L. and Lynch,  Galen F. and Aronov,  Dmitriy},
  year = {2021},
  month = jul,
  pages = {343–348}
}

@article{Mantel1967,
    author = {Mantel, Nathan},
    title = {The Detection of Disease Clustering and a Generalized Regression Approach},
    journal = {Cancer Research},
    volume = {27},
    number = {2\_Part\_1},
    pages = {209-220},
    year = {1967},
    month = {02},
    issn = {0008-5472},
    eprint = {https://aacrjournals.org/cancerres/article-pdf/27/2_Part_1/209/2382183/cr0272p10209.pdf}
}

@article{Mantel1970,
  title = {A Technique of Nonparametric Multivariate Analysis},
  volume = {26},
  ISSN = {0006-341X},
  url = {http://dx.doi.org/10.2307/2529108},
  DOI = {10.2307/2529108},
  number = {3},
  journal = {Biometrics},
  publisher = {JSTOR},
  author = {Mantel,  Nathan and Valand,  Ranchhodbhai S.},
  year = {1970},
  month = sep,
  pages = {547}
}

@article{Kriegeskorte2008,
  title = {Representational similarity analysis – connecting the branches of systems neuroscience},
  DOI = {10.3389/neuro.06.004.2008},
  journal = {Frontiers in Systems Neuroscience},
  author = {Kriegeskorte,  Nikolaus and Mur, Marieke and Bandettini, Peter},
  ISSN = {1662-5137},
  year = {2008}
}

@article{Squire1992,
  title = {Memory and the hippocampus: A synthesis from findings with rats,  monkeys,  and humans.},
  volume = {99},
  ISSN = {0033-295X},
  url = {http://dx.doi.org/10.1037/0033-295X.99.2.195},
  DOI = {10.1037/0033-295x.99.2.195},
  number = {2},
  journal = {Psychological Review},
  publisher = {American Psychological Association (APA)},
  author = {Squire,  Larry R.},
  year = {1992},
  pages = {195–231}
}

@article{Bailey1996,
  title = {Toward a molecular definition of long-term memory storage},
  volume = {93},
  ISSN = {1091-6490},
  url = {http://dx.doi.org/10.1073/pnas.93.24.13445},
  DOI = {10.1073/pnas.93.24.13445},
  number = {24},
  journal = {Proceedings of the National Academy of Sciences},
  publisher = {Proceedings of the National Academy of Sciences},
  author = {Bailey,  Craig H. and Bartsch,  Dusan and Kandel,  Eric R.},
  year = {1996},
  month = nov,
  pages = {13445–13452}
}

@article{Thome2017,
  title = {Evidence for an Evolutionarily Conserved Memory Coding Scheme in the Mammalian Hippocampus},
  volume = {37},
  ISSN = {1529-2401},
  url = {http://dx.doi.org/10.1523/JNEUROSCI.3057-16.2017},
  DOI = {10.1523/jneurosci.3057-16.2017},
  number = {10},
  journal = {The Journal of Neuroscience},
  publisher = {Society for Neuroscience},
  author = {Thome,  Alexander and Marrone,  Diano F. and Ellmore,  Timothy M. and Chawla,  Monica K. and Lipa,  Peter and Ramirez-Amaya,  Victor and Lisanby,  Sarah H. and McNaughton,  Bruce L. and Barnes,  Carol A.},
  year = {2017},
  month = feb,
  pages = {2795–2801}
}

@article{Jacobs2003,
  title = {The Evolution of the Cognitive Map},
  volume = {62},
  ISSN = {1421-9743},
  url = {http://dx.doi.org/10.1159/000072443},
  DOI = {10.1159/000072443},
  number = {2},
  journal = {Brain,  Behavior and Evolution},
  publisher = {S. Karger AG},
  author = {Jacobs,  Lucia F.},
  year = {2003},
  pages = {128–139}
}

@article{Roxin2013,
  title = {{Efficient Partitioning of Memory Systems and Its Importance for Memory Consolidation}},
  volume = {9},
  ISSN = {1553-7358},
  url = {http://dx.doi.org/10.1371/journal.pcbi.1003146},
  DOI = {10.1371/journal.pcbi.1003146},
  number = {7},
  journal = {PLoS Computational Biology},
  publisher = {Public Library of Science (PLoS)},
  author = {Roxin,  Alex and Fusi,  Stefano},
  editor = {Beck,  Jeff},
  year = {2013},
  month = jul,
  pages = {e1003146}
}

@article{Sherry1989_behav,
  title = {Hippocampus and memory for food caches in black-capped chickadees.},
  volume = {103},
  ISSN = {0735-7044},
  url = {http://dx.doi.org/10.1037/0735-7044.103.2.308},
  DOI = {10.1037/0735-7044.103.2.308},
  number = {2},
  journal = {Behavioral Neuroscience},
  publisher = {American Psychological Association (APA)},
  author = {Sherry,  David F. and Vaccarino,  Anthony L.},
  year = {1989},
  month = apr,
  pages = {308–318}
}

@article{Sherry1989_brain,
  title = {{The Hippocampal Complex of Food-Storing Birds}},
  volume = {34},
  ISSN = {1421-9743},
  url = {http://dx.doi.org/10.1159/000116516},
  DOI = {10.1159/000116516},
  number = {5},
  journal = {Brain,  Behavior and Evolution},
  publisher = {S. Karger AG},
  author = {Sherry,  David F. and Vaccarino,  Anthony L. and Buckenham,  Karen and Herz,  Rachel S.},
  year = {1989},
  pages = {308–317}
}

@article{Garamszegi2005,
  title = {Continental variation in relative hippocampal volume in birds: the phylogenetic extent of the effect and the potential role of winter temperatures},
  volume = {1},
  ISSN = {1744-957X},
  url = {http://dx.doi.org/10.1098/rsbl.2005.0328},
  DOI = {10.1098/rsbl.2005.0328},
  number = {3},
  journal = {Biology Letters},
  publisher = {The Royal Society},
  author = {Garamszegi,  László Zsolt and Lucas,  Jeffrey R},
  year = {2005},
  month = jun,
  pages = {330–333}
}

@article{Krebs1989,
  title = {Hippocampal specialization of food-storing birds.},
  volume = {86},
  ISSN = {1091-6490},
  url = {http://dx.doi.org/10.1073/pnas.86.4.1388},
  DOI = {10.1073/pnas.86.4.1388},
  number = {4},
  journal = {Proceedings of the National Academy of Sciences},
  publisher = {Proceedings of the National Academy of Sciences},
  author = {Krebs,  J R and Sherry,  D F and Healy,  S D and Perry,  V H and Vaccarino,  A L},
  year = {1989},
  month = feb,
  pages = {1388–1392}
}

@article{Smulders2017,
  title = {{The Avian Hippocampal Formation and the Stress Response}},
  volume = {90},
  ISSN = {1421-9743},
  url = {http://dx.doi.org/10.1159/000477654},
  DOI = {10.1159/000477654},
  number = {1},
  journal = {Brain,  Behavior and Evolution},
  publisher = {S. Karger AG},
  author = {Smulders,  Tom V.},
  year = {2017},
  pages = {81–91}
}

@article{Pravosudov2015,
  title = {Environmental Influences on Spatial Memory and the Hippocampus in Food-Caching Chickadees},
  volume = {10},
  ISSN = {1911-4745},
  url = {http://dx.doi.org/10.3819/ccbr.2015.100002},
  DOI = {10.3819/ccbr.2015.100002},
  journal = {Comparative Cognition \& Behavior Reviews},
  publisher = {Comparative Cognition Society},
  author = {Pravosudov,  Vladimir V. and Roth,  Timothy C. and LaDage,  Lara D. and Freas,  Cody A.},
  year = {2015},
  pages = {25–43}
}

@article{Aronov2017,
  title = {Mapping of a non-spatial dimension by the hippocampal–entorhinal circuit},
  volume = {543},
  ISSN = {1476-4687},
  url = {http://dx.doi.org/10.1038/nature21692},
  DOI = {10.1038/nature21692},
  number = {7647},
  journal = {Nature},
  publisher = {Springer Science and Business Media LLC},
  author = {Aronov,  Dmitriy and Nevers,  Rhino and Tank,  David W.},
  year = {2017},
  month = mar,
  pages = {719–722}
}

@article{Clayton1995,
  title = {Memory in food-storing birds: from behaviour to brain},
  volume = {5},
  ISSN = {0959-4388},
  url = {http://dx.doi.org/10.1016/0959-4388(95)80020-4},
  DOI = {10.1016/0959-4388(95)80020-4},
  number = {2},
  journal = {Current Opinion in Neurobiology},
  publisher = {Elsevier BV},
  author = {Clayton,  Nicky S and Krebs,  John R},
  year = {1995},
  month = apr,
  pages = {149–154}
}

@article{Brodbeck1995,
  title = {Matching location and color of a compound stimulus: Comparison of a food-storing and a nonstoring bird species.},
  volume = {21},
  ISSN = {0097-7403},
  url = {http://dx.doi.org/10.1037/0097-7403.21.1.64},
  DOI = {10.1037/0097-7403.21.1.64},
  number = {1},
  journal = {Journal of Experimental Psychology: Animal Behavior Processes},
  publisher = {American Psychological Association (APA)},
  author = {Brodbeck,  David R. and Shettleworth,  Sara J.},
  year = {1995},
  month = jan,
  pages = {64–77}
}

@article{Balda1992,
  title = {Long-term spatial memory in clark’s nutcracker,  Nucifraga columbiana},
  volume = {44},
  ISSN = {0003-3472},
  url = {http://dx.doi.org/10.1016/S0003-3472(05)80302-1},
  DOI = {10.1016/s0003-3472(05)80302-1},
  number = {4},
  journal = {Animal Behaviour},
  publisher = {Elsevier BV},
  author = {Balda,  Russell P. and Kamil,  Alan C.},
  year = {1992},
  month = oct,
  pages = {761–769}
}

@article{Thompson1990,
  title = {Long-term stability of the place-field activity of single units recorded from the dorsal hippocampus of freely behaving rats},
  volume = {509},
  ISSN = {0006-8993},
  url = {http://dx.doi.org/10.1016/0006-8993(90)90555-P},
  DOI = {10.1016/0006-8993(90)90555-p},
  number = {2},
  journal = {Brain Research},
  publisher = {Elsevier BV},
  author = {Thompson,  L.T. and Best,  P.J.},
  year = {1990},
  month = feb,
  pages = {299–308}
}

@article{Redish2001,
  title = {{Independence of Firing Correlates of Anatomically Proximate Hippocampal Pyramidal Cells}},
  volume = {21},
  ISSN = {1529-2401},
  url = {http://dx.doi.org/10.1523/JNEUROSCI.21-05-j0004.2001},
  DOI = {10.1523/jneurosci.21-05-j0004.2001},
  number = {5},
  journal = {The Journal of Neuroscience},
  publisher = {Society for Neuroscience},
  author = {Redish,  A. D. and Battaglia,  F. P. and Chawla,  M. K. and Ekstrom,  A. D. and Gerrard,  J. L. and Lipa,  P. and Rosenzweig,  E. S. and Worley,  P. F. and Guzowski,  J. F. and McNaughton,  B. L. and Barnes,  C. A.},
  year = {2001},
  month = mar,
  pages = {RC134–RC134}
}

@article{Waydo2006,
  title = {{Sparse Representation in the Human Medial Temporal Lobe}},
  volume = {26},
  ISSN = {1529-2401},
  url = {http://dx.doi.org/10.1523/JNEUROSCI.2101-06.2006},
  DOI = {10.1523/jneurosci.2101-06.2006},
  number = {40},
  journal = {The Journal of Neuroscience},
  publisher = {Society for Neuroscience},
  author = {Waydo,  Stephen and Kraskov,  Alexander and Quian Quiroga,  Rodrigo and Fried,  Itzhak and Koch,  Christof},
  year = {2006},
  month = oct,
  pages = {10232–10234}
}

@article{French1999,
  title = {Catastrophic forgetting in connectionist networks},
  volume = {3},
  ISSN = {1364-6613},
  url = {http://dx.doi.org/10.1016/S1364-6613(99)01294-2},
  DOI = {10.1016/s1364-6613(99)01294-2},
  number = {4},
  journal = {Trends in Cognitive Sciences},
  publisher = {Elsevier BV},
  author = {French,  R},
  year = {1999},
  month = apr,
  pages = {128–135}
}

@book{Dayan2005,
  author    = {Dayan, Peter and Abbott, L. F.},
  title     = {{Theoretical Neuroscience: Computational and Mathematical
               Modeling of Neural Systems}},
  publisher = {MIT Press},
  address   = {Cambridge, MA},
  year      = {2005},
  isbn      = {9780262541855},
  series    = {Computational Neuroscience Series}
}

@article{Valiant2005,
  title = {{Memorization and Association on a Realistic Neural Model}},
  volume = {17},
  ISSN = {1530-888X},
  url = {http://dx.doi.org/10.1162/0899766053019890},
  DOI = {10.1162/0899766053019890},
  number = {3},
  journal = {Neural Computation},
  publisher = {MIT Press},
  author = {Valiant,  Leslie G.},
  year = {2005},
  month = mar,
  pages = {527–555}
}

@article{Cunningham2014,
  title = {Dimensionality reduction for large-scale neural recordings},
  volume = {17},
  ISSN = {1546-1726},
  url = {http://dx.doi.org/10.1038/nn.3776},
  DOI = {10.1038/nn.3776},
  number = {11},
  journal = {Nature Neuroscience},
  publisher = {Springer Science and Business Media LLC},
  author = {Cunningham,  John P and Yu,  Byron M},
  year = {2014},
  month = aug,
  pages = {1500–1509}
}

@article{Burak2009,
  title = {{Accurate Path Integration in Continuous Attractor Network Models of Grid Cells}},
  volume = {5},
  ISSN = {1553-7358},
  url = {http://dx.doi.org/10.1371/journal.pcbi.1000291},
  DOI = {10.1371/journal.pcbi.1000291},
  number = {2},
  journal = {PLoS Computational Biology},
  publisher = {Public Library of Science (PLoS)},
  author = {Burak,  Yoram and Fiete,  Ila R.},
  editor = {Sporns,  Olaf},
  year = {2009},
  month = feb,
  pages = {e1000291}
}

@article{Fusi2016,
  title = {Why neurons mix: high dimensionality for higher cognition},
  volume = {37},
  ISSN = {0959-4388},
  url = {http://dx.doi.org/10.1016/j.conb.2016.01.010},
  DOI = {10.1016/j.conb.2016.01.010},
  journal = {Current Opinion in Neurobiology},
  publisher = {Elsevier BV},
  author = {Fusi,  Stefano and Miller,  Earl K and Rigotti,  Mattia},
  year = {2016},
  month = apr,
  pages = {66–74}
}

@article{Rigotti2013,
  title = {The importance of mixed selectivity in complex cognitive tasks},
  volume = {497},
  ISSN = {1476-4687},
  url = {http://dx.doi.org/10.1038/nature12160},
  DOI = {10.1038/nature12160},
  number = {7451},
  journal = {Nature},
  publisher = {Springer Science and Business Media LLC},
  author = {Rigotti,  Mattia and Barak,  Omri and Warden,  Melissa R. and Wang,  Xiao-Jing and Daw,  Nathaniel D. and Miller,  Earl K. and Fusi,  Stefano},
  year = {2013},
  month = may,
  pages = {585–590}
}

@article{Jazayeri2021,
  title = {Interpreting neural computations by examining intrinsic and embedding dimensionality of neural activity},
  volume = {70},
  ISSN = {0959-4388},
  url = {http://dx.doi.org/10.1016/j.conb.2021.08.002},
  DOI = {10.1016/j.conb.2021.08.002},
  journal = {Current Opinion in Neurobiology},
  publisher = {Elsevier BV},
  author = {Jazayeri,  Mehrdad and Ostojic,  Srdjan},
  year = {2021},
  month = oct,
  pages = {113–120}
}

@article{Saxe2020,
  title = {If deep learning is the answer,  what is the question?},
  volume = {22},
  ISSN = {1471-0048},
  url = {http://dx.doi.org/10.1038/s41583-020-00395-8},
  DOI = {10.1038/s41583-020-00395-8},
  number = {1},
  journal = {Nature Reviews Neuroscience},
  publisher = {Springer Science and Business Media LLC},
  author = {Saxe,  Andrew and Nelli,  Stephanie and Summerfield,  Christopher},
  year = {2020},
  month = nov,
  pages = {55–67}
}

@book{Brahms1886S4,
  author    = {Brahms, Johannes},
  title     = {{Symphonie Nr. 4 in e-Moll, Op. 98}},
  publisher = {N. Simrock},
  address   = {Berlin},
  year      = {1886},
  note      = {First edition. Plate 8686. Engraved by C.G. Röder, Leipzig.},
}

@article{Knapp1989,
  title = {{The Finale of Brahms’s Fourth Symphony: The Tale of the Subject}},
  volume = {13},
  ISSN = {0148-2076},
  url = {http://dx.doi.org/10.2307/746207},
  DOI = {10.2307/746207},
  number = {1},
  journal = {19th-Century Music},
  publisher = {University of California Press},
  author = {Knapp,  Raymond},
  year = {1989},
  pages = {3–17}
}

@article{Pascall1989,
  title = {{Genre and the Finale of Brahms’s Fourth Symphony}},
  volume = {8},
  ISSN = {0262-5245},
  url = {http://dx.doi.org/10.2307/854289},
  DOI = {10.2307/854289},
  number = {3},
  journal = {Music Analysis},
  publisher = {JSTOR},
  author = {Pascall,  Robert},
  year = {1989},
  month = oct,
  pages = {233}
}

@article{Kandel2001,
  title     = {The molecular biology of memory storage: a dialogue between
               genes and synapses},
  author    = {Kandel, Eric R.},
  journal   = {Science},
  publisher = {American Association for the Advancement of Science (AAAS)},
  volume    =  {294},
  number    =  {5544},
  pages     = {1030--1038},
  month     =  nov,
  year      =  {2001}
}

@article{Benna2016,
  title = {Computational principles of synaptic memory consolidation},
  volume = {19},
  ISSN = {1546-1726},
  url = {http://dx.doi.org/10.1038/nn.4401},
  DOI = {10.1038/nn.4401},
  number = {12},
  journal = {Nature Neuroscience},
  publisher = {Springer Science and Business Media LLC},
  author = {Benna,  Marcus K and Fusi,  Stefano},
  year = {2016},
  month = oct,
  pages = {1697–1706}
}

@article{Bernardi2020,
  title = {{The Geometry of Abstraction in the Hippocampus and Prefrontal Cortex}},
  volume = {183},
  ISSN = {0092-8674},
  url = {http://dx.doi.org/10.1016/j.cell.2020.09.031},
  DOI = {10.1016/j.cell.2020.09.031},
  number = {4},
  journal = {Cell},
  publisher = {Elsevier BV},
  author = {Bernardi,  Silvia and Benna,  Marcus K. and Rigotti,  Mattia and Munuera,  Jér\^ome and Fusi,  Stefano and Salzman,  C. Daniel},
  year = {2020},
  month = nov,
  pages = {954--967.e21}
}

@article{Benna2021,
  title = {Place cells may simply be memory cells: Memory compression leads to spatial tuning and history dependence},
  volume = {118},
  ISSN = {1091-6490},
  url = {http://dx.doi.org/10.1073/pnas.2018422118},
  DOI = {10.1073/pnas.2018422118},
  number = {51},
  journal = {Proceedings of the National Academy of Sciences},
  publisher = {Proceedings of the National Academy of Sciences},
  author = {Benna,  Marcus K. and Fusi,  Stefano},
  year = {2021},
  month = dec 
}

@article{Boyle2024,
  title = {Tuned geometries of hippocampal representations meet the computational demands of social memory},
  volume = {112},
  ISSN = {0896-6273},
  url = {http://dx.doi.org/10.1016/j.neuron.2024.01.021},
  DOI = {10.1016/j.neuron.2024.01.021},
  number = {8},
  journal = {Neuron},
  publisher = {Elsevier BV},
  author = {Boyle,  Lara M. and Posani,  Lorenzo and Irfan,  Sarah and Siegelbaum,  Steven A. and Fusi,  Stefano},
  year = {2024},
  month = apr,
  pages = {1358--1371.e9}
}

@article{Fusi2007,
  title = {Limits on the memory storage capacity of bounded synapses},
  volume = {10},
  ISSN = {1546-1726},
  url = {http://dx.doi.org/10.1038/nn1859},
  DOI = {10.1038/nn1859},
  number = {4},
  journal = {Nature Neuroscience},
  publisher = {Springer Science and Business Media LLC},
  author = {Fusi,  Stefano and Abbott,  L F},
  year = {2007},
  month = mar,
  pages = {485–493}
}

@article{Gulli2019,
  title = {Context-dependent representations of objects and space in the primate hippocampus during virtual navigation},
  volume = {23},
  ISSN = {1546-1726},
  url = {http://dx.doi.org/10.1038/s41593-019-0548-3},
  DOI = {10.1038/s41593-019-0548-3},
  number = {1},
  journal = {Nature Neuroscience},
  publisher = {Springer Science and Business Media LLC},
  author = {Gulli,  Roberto A. and Duong,  Lyndon R. and Corrigan,  Benjamin W. and Doucet,  Guillaume and Williams,  Sylvain and Fusi,  Stefano and Martinez-Trujillo,  Julio C.},
  year = {2019},
  month = dec,
  pages = {103–112}
}

@article{Courellis2024,
  title = {Abstract representations emerge in human hippocampal neurons during inference},
  volume = {632},
  ISSN = {1476-4687},
  url = {http://dx.doi.org/10.1038/s41586-024-07799-x},
  DOI = {10.1038/s41586-024-07799-x},
  number = {8026},
  journal = {Nature},
  publisher = {Springer Science and Business Media LLC},
  author = {Courellis,  Hristos S. and Minxha,  Juri and Cardenas,  Araceli R. and Kimmel,  Daniel L. and Reed,  Chrystal M. and Valiante,  Taufik A. and Salzman,  C. Daniel and Mamelak,  Adam N. and Fusi,  Stefano and Rutishauser,  Ueli},
  year = {2024},
  month = aug,
  pages = {841–849}
}

@article{Hopfield1982,
  title = {Neural networks and physical systems with emergent collective computational abilities.},
  volume = {79},
  ISSN = {1091-6490},
  url = {http://dx.doi.org/10.1073/pnas.79.8.2554},
  DOI = {10.1073/pnas.79.8.2554},
  number = {8},
  journal = {Proceedings of the National Academy of Sciences},
  publisher = {Proceedings of the National Academy of Sciences},
  author = {Hopfield,  J J},
  year = {1982},
  month = apr,
  pages = {2554–2558}
}

@article{Amit1987,
  title = {Statistical mechanics of neural networks near saturation},
  volume = {173},
  ISSN = {0003-4916},
  url = {http://dx.doi.org/10.1016/0003-4916(87)90092-3},
  DOI = {10.1016/0003-4916(87)90092-3},
  number = {1},
  journal = {Annals of Physics},
  publisher = {Elsevier BV},
  author = {Amit,  Daniel J and Gutfreund,  Hanoch and Sompolinsky,  H},
  year = {1987},
  month = jan,
  pages = {30–67}
}

@inbook{McCloskey1989,
  title = {{Catastrophic Interference in Connectionist Networks: The Sequential Learning Problem}},
  ISBN = {9780125433242},
  ISSN = {0079-7421},
  url = {http://dx.doi.org/10.1016/S0079-7421(08)60536-8},
  DOI = {10.1016/s0079-7421(08)60536-8},
  booktitle = {Psychology of Learning and Motivation},
  publisher = {Elsevier},
  author = {McCloskey,  Michael and Cohen,  Neal J.},
  year = {1989},
  pages = {109–165}
}

@book{Rolls1998NeuralNA,
  title={Neural networks and brain function},
  author={Edmund T. Rolls and Alessandro Treves},
  year={1998},
  publisher = {Oxford University Press},
  address   = {Oxford},
  doi       = {10.1093/acprof:oso/9780198524328.001.0001},
  url={https://api.semanticscholar.org/CorpusID:14588175}
}

@article{Chaudhuri2019,
  title = {The intrinsic attractor manifold and population dynamics of a canonical cognitive circuit across waking and sleep},
  volume = {22},
  ISSN = {1546-1726},
  url = {http://dx.doi.org/10.1038/s41593-019-0460-x},
  DOI = {10.1038/s41593-019-0460-x},
  number = {9},
  journal = {Nature Neuroscience},
  publisher = {Springer Science and Business Media LLC},
  author = {Chaudhuri,  Rishidev and Ger\c{c}ek,  Berk and Pandey,  Biraj and Peyrache,  Adrien and Fiete,  Ila},
  year = {2019},
  month = aug,
  pages = {1512–1520}
}

@article{Renart2010,
  title = {{The Asynchronous State in Cortical Circuits}},
  volume = {327},
  ISSN = {1095-9203},
  url = {http://dx.doi.org/10.1126/science.1179850},
  DOI = {10.1126/science.1179850},
  number = {5965},
  journal = {Science},
  publisher = {American Association for the Advancement of Science (AAAS)},
  author = {Renart,  Alfonso and de la Rocha,  Jaime and Bartho,  Peter and Hollender,  Liad and Parga,  Néstor and Reyes,  Alex and Harris,  Kenneth D.},
  year = {2010},
  month = jan,
  pages = {587–590}
}

@article{Canning1988,
  title = {Partially connected models of neural networks},
  volume = {21},
  ISSN = {1361-6447},
  url = {http://dx.doi.org/10.1088/0305-4470/21/15/016},
  DOI = {10.1088/0305-4470/21/15/016},
  number = {15},
  journal = {Journal of Physics A: Mathematical and General},
  publisher = {IOP Publishing},
  author = {Canning,  A and Gardner,  E},
  year = {1988},
  month = aug,
  pages = {3275–3284}
}

@article{Compte2000,
  title = {{Synaptic Mechanisms and Network Dynamics Underlying Spatial Working Memory in a Cortical Network Model}},
  volume = {10},
  ISSN = {1460-2199},
  url = {http://dx.doi.org/10.1093/cercor/10.9.910},
  DOI = {10.1093/cercor/10.9.910},
  number = {9},
  journal = {Cerebral Cortex},
  publisher = {Oxford University Press (OUP)},
  author = {Compte,  A.},
  year = {2000},
  month = sep,
  pages = {910–923}
}

@article{Burak2012,
  title = {Fundamental limits on persistent activity in networks of noisy neurons},
  volume = {109},
  ISSN = {1091-6490},
  url = {http://dx.doi.org/10.1073/pnas.1117386109},
  DOI = {10.1073/pnas.1117386109},
  number = {43},
  journal = {Proceedings of the National Academy of Sciences},
  publisher = {Proceedings of the National Academy of Sciences},
  author = {Burak,  Yoram and Fiete,  Ila R.},
  year = {2012},
  month = oct,
  pages = {17645–17650}
}

@book{Buzsaki2006,
  author    = {Buzs{\'a}ki, Gy{\"o}rgy},
  title     = {Rhythms of the Brain},
  publisher = {Oxford University Press},
  year      = {2006},
  address   = {Oxford},
  isbn      = {9780199828234},
  doi       = {10.1093/acprof:oso/9780195301069.001.0001}
}

@article{Langdon2023,
  title = {A unifying perspective on neural manifolds and circuits for cognition},
  volume = {24},
  ISSN = {1471-0048},
  url = {http://dx.doi.org/10.1038/s41583-023-00693-x},
  DOI = {10.1038/s41583-023-00693-x},
  number = {6},
  journal = {Nature Reviews Neuroscience},
  publisher = {Springer Science and Business Media LLC},
  author = {Langdon,  Christopher and Genkin,  Mikhail and Engel,  Tatiana A.},
  year = {2023},
  month = apr,
  pages = {363–377}
}

@article{Sorscher2023,
  title = {A unified theory for the computational and mechanistic origins of grid cells},
  volume = {111},
  ISSN = {0896-6273},
  url = {http://dx.doi.org/10.1016/j.neuron.2022.10.003},
  DOI = {10.1016/j.neuron.2022.10.003},
  number = {1},
  journal = {Neuron},
  publisher = {Elsevier BV},
  author = {Sorscher,  Ben and Mel,  Gabriel C. and Ocko,  Samuel A. and Giocomo,  Lisa M. and Ganguli,  Surya},
  year = {2023},
  month = jan,
  pages = {121--137.e13}
}

@article{Raju2024,
  title = {Space is a latent sequence: A theory of the hippocampus},
  volume = {10},
  ISSN = {2375-2548},
  url = {http://dx.doi.org/10.1126/sciadv.adm8470},
  DOI = {10.1126/sciadv.adm8470},
  number = {31},
  journal = {Science Advances},
  publisher = {American Association for the Advancement of Science (AAAS)},
  author = {Raju,  Rajkumar Vasudeva and Guntupalli,  J. Swaroop and Zhou,  Guangyao and Wendelken,  Carter and Lázaro-Gredilla,  Miguel and George,  Dileep},
  year = {2024},
  month = aug 
}

@article{Mallory2018,
  title = {Grid scale drives the scale and long-term stability of place maps},
  volume = {21},
  ISSN = {1546-1726},
  url = {http://dx.doi.org/10.1038/s41593-017-0055-3},
  DOI = {10.1038/s41593-017-0055-3},
  number = {2},
  journal = {Nature Neuroscience},
  publisher = {Springer Science and Business Media LLC},
  author = {Mallory,  Caitlin S. and Hardcastle,  Kiah and Bant,  Jason S. and Giocomo,  Lisa M.},
  year = {2018},
  month = jan,
  pages = {270–282}
}

@article{Hardcastle2015,
  title = {{Environmental Boundaries as an Error Correction Mechanism for Grid Cells}},
  volume = {86},
  ISSN = {0896-6273},
  url = {http://dx.doi.org/10.1016/j.neuron.2015.03.039},
  DOI = {10.1016/j.neuron.2015.03.039},
  number = {3},
  journal = {Neuron},
  publisher = {Elsevier BV},
  author = {Hardcastle,  Kiah and Ganguli,  Surya and Giocomo,  Lisa M.},
  year = {2015},
  month = may,
  pages = {827–839}
}

@article{MeissnerBernard2024,
  title = {Geometry and dynamics of representations in a precisely balanced memory network related to olfactory cortex},
  url = {http://dx.doi.org/10.7554/eLife.96303.1},
  DOI = {10.7554/elife.96303.1},
  journal = {eLife},
  publisher = {eLife Sciences Publications,  Ltd},
  author = {Meissner-Bernard,  Claire and Zenke,  Friedemann and Friedrich,  Rainer W.},
  year = {2024},
  month = may 
}

@article{Denve2016,
  title = {Efficient codes and balanced networks},
  volume = {19},
  ISSN = {1546-1726},
  url = {http://dx.doi.org/10.1038/nn.4243},
  DOI = {10.1038/nn.4243},
  number = {3},
  journal = {Nature Neuroscience},
  publisher = {Springer Science and Business Media LLC},
  author = {Denève,  Sophie and Machens,  Christian K},
  year = {2016},
  month = feb,
  pages = {375–382}
}

@article{Morales2025,
  title = {Representational drift and learning-induced stabilization in the piriform cortex},
  volume = {122},
  ISSN = {1091-6490},
  url = {http://dx.doi.org/10.1073/pnas.2501811122},
  DOI = {10.1073/pnas.2501811122},
  number = {29},
  journal = {Proceedings of the National Academy of Sciences},
  publisher = {Proceedings of the National Academy of Sciences},
  author = {Morales,  Guillermo B. and Muñoz,  Miguel A. and Tu,  Yuhai},
  year = {2025},
  month = jul 
}

@article{Marr1971,
  title = {Simple memory: a theory for archicortex},
  volume = {262},
  ISSN = {2054-0280},
  url = {http://dx.doi.org/10.1098/rstb.1971.0078},
  DOI = {10.1098/rstb.1971.0078},
  number = {841},
  journal = {Philosophical Transactions of the Royal Society of London. B,  Biological Sciences},
  publisher = {The Royal Society},
  author = {Marr,  D.},
  year = {1971},
  month = jul,
  pages = {23–81}
}

@article{Treves1994,
  title = {Computational analysis of the role of the hippocampus in memory},
  volume = {4},
  ISSN = {1098-1063},
  url = {http://dx.doi.org/10.1002/hipo.450040319},
  DOI = {10.1002/hipo.450040319},
  number = {3},
  journal = {Hippocampus},
  publisher = {Wiley},
  author = {Treves,  Alessandro and Rolls,  Edmund T.},
  year = {1994},
  month = jun,
  pages = {374–391}
}

@inbook{Fusi2024,
  title = {{Memory Capacity of Neural Network Models}},
  ISBN = {9780190918019},
  url = {http://dx.doi.org/10.1093/oxfordhb/9780190917982.013.26},
  DOI = {10.1093/oxfordhb/9780190917982.013.26},
  booktitle = {The Oxford Handbook of Human Memory,  Two Volume Pack},
  publisher = {Oxford University Press},
  author = {Fusi,  Stefano},
  year = {2024},
  month = jul,
  pages = {740–764}
}

@article{Bakker2008,
  title = {{Pattern Separation in the Human Hippocampal CA3 and Dentate Gyrus}},
  volume = {319},
  ISSN = {1095-9203},
  url = {http://dx.doi.org/10.1126/science.1152882},
  DOI = {10.1126/science.1152882},
  number = {5870},
  journal = {Science},
  publisher = {American Association for the Advancement of Science (AAAS)},
  author = {Bakker,  Arnold and Kirwan,  C. Brock and Miller,  Michael and Stark,  Craig E. L.},
  year = {2008},
  month = mar,
  pages = {1640–1642}
}

@article{Colgin2008,
  title = {Understanding memory through hippocampal remapping},
  volume = {31},
  ISSN = {0166-2236},
  url = {http://dx.doi.org/10.1016/j.tins.2008.06.008},
  DOI = {10.1016/j.tins.2008.06.008},
  number = {9},
  journal = {Trends in Neurosciences},
  publisher = {Elsevier BV},
  author = {Colgin,  Laura Lee and Moser,  Edvard I. and Moser,  May-Britt},
  year = {2008},
  month = sep,
  pages = {469–477}
}

@article{McNaughton1987,
  title = {Hippocampal synaptic enhancement and information storage within a distributed memory system},
  volume = {10},
  ISSN = {0166-2236},
  url = {http://dx.doi.org/10.1016/0166-2236(87)90011-7},
  DOI = {10.1016/0166-2236(87)90011-7},
  number = {10},
  journal = {Trends in Neurosciences},
  publisher = {Elsevier BV},
  author = {McNaughton,  B.L. and Morris,  R.G.M.},
  year = {1987},
  month = jan,
  pages = {408–415}
}

@book{Kandel2006book,
  title     = {In search of memory},
  author    = {Kandel, Eric R.},
  publisher = {WW Norton},
  month     =  mar,
  year      =  {2007},
  address   = {New York, NY},
  isbn = {978-0393329377}
}

@book{Kandel2016,
  title     = {Reductionism in art and brain science},
  author    = {Kandel, Eric R.},
  publisher = {Columbia University Press},
  month     =  aug,
  year      =  {2016},
  address   = {New York, NY},
  isbn = {978-0231179621}
}

@article{Vogels2011,
  title = {{Inhibitory Plasticity Balances Excitation and Inhibition in Sensory Pathways and Memory Networks}},
  volume = {334},
  ISSN = {1095-9203},
  url = {http://dx.doi.org/10.1126/science.1211095},
  DOI = {10.1126/science.1211095},
  number = {6062},
  journal = {Science},
  publisher = {American Association for the Advancement of Science (AAAS)},
  author = {Vogels,  T. P. and Sprekeler,  H. and Zenke,  F. and Clopath,  C. and Gerstner,  W.},
  year = {2011},
  month = dec,
  pages = {1569–1573}
}

\newpage

\appendix
\setcounter{figure}{0}

\renewcommand{\figurename}{Figure}

\renewcommand{\thefigure}{S\arabic{figure}}
\setcounter{table}{0}

\renewcommand{\tablename}{Table}

\renewcommand{\thetable}{S\arabic{table}}
\section{Methods}
\label{app:methods}


\subsection{Subjects and electrophysiological recordings}

All analyses were performed on publicly available \textit{in vivo} extracellular electrophysiological recordings from the avian hippocampal formation (dorsomedial pallium)~\citep{Payne2021, Payne2021data}. The dataset comprised 755 single units recorded from food-caching black-capped chickadees (\textit{Poecile atricapillus}; 39 sessions across 9 birds) and 238 single units from non-caching zebra finches (\textit{Taeniopygia guttata}; 8 sessions across 9 birds), for a total of 993 units. Data were loaded from MATLAB v5 binary files (\texttt{RESULTS\_T.mat}, \texttt{RESULTS\_Z.mat}) and parsed using a custom recursive miMATRIX reader (for MCOS table objects embedded in the MAT v5 subsystem) or, where possible, using \texttt{scipy.io.loadmat}. Behavioral telemetry (position, head angle, movement speed) was extracted from the accompanying \texttt{B} (behavior) variable stored as a MATLAB struct.

\subsection{Cell-type classification}

Putative excitatory (E) and inhibitory (I) neurons were classified using Ward hierarchical agglomerative clustering on standardized waveform features. For each unit, two features were extracted: (i) spike width (peak-to-trough duration, converted to milliseconds) and (ii) peak-to-peak ratio (\texttt{pp\_ratio}). Features were $z$-scored prior to clustering:
\begin{equation*}
\mathbf{f}_{\text{std}} = \frac{\mathbf{f} - \bar{\mathbf{f}}}{\sigma_{\mathbf{f}}}
\end{equation*}
Ward linkage was applied to the standardized feature matrix and the dendrogram was cut at $k = 2$ clusters. The cluster with the lower mean spike width was assigned as narrow-spiking (putative inhibitory) and the remaining cluster as broad-spiking (putative excitatory). This procedure yielded a classification concordance of $\sim$99\% with the published labels (536/219 versus reported 538/217 for chickadees). Spatial selectivity was assessed by computing the spatial information (\texttt{info}) of each unit and comparing it against a null distribution of 200 bin-shuffled surrogates; units with information exceeding the 99th percentile of the shuffle distribution ($p < 0.01$) were classified as spatially selective.

\subsection{Construction of population rate maps}

For each recording session, two-dimensional arena occupancy was discretized into a $40 \times 40$ grid ($S = 1{,}600$ spatial bins). Firing rate maps were computed for each neuron as the spike count in each bin divided by the occupancy time, yielding a one-dimensional vector $\mathbf{r}_k \in \mathbb{R}^{S}$ per neuron $k$. Bins with no occupancy were retained as \texttt{NaN} rather than zero-filled; all pairwise distance computations used pairwise deletion (see below). For place field detection, which requires spatially continuous maps, \texttt{NaN} bins were locally replaced with zero prior to Gaussian smoothing. The population response matrix $\mathbf{M} \in \mathbb{R}^{N \times S}$ was constructed by stacking the rate maps of all $N$ neurons recorded in a given session, where each row represents a neuron and each column represents a spatial bin. Sessions were included in all subsequent analyses only if they contained at least $N_{\min} \geq 5$ excitatory neurons; for the E/I analyses (Tier~2), both cell types required a minimum of $N_{\min} = 3$ neurons per type. Rate maps were $z$-scored neuron-wise prior to all geometric analyses:
\begin{equation*}
\tilde{r}_{k,s} = \frac{r_{k,s} - \bar{r}_k}{\sigma_{r_k}}, \quad \sigma_{r_k} > 0
\end{equation*}
Neurons with zero variance ($\sigma_{r_k} = 0$) were assigned unit standard deviation to avoid division by zero. The per-neuron mean $\bar{r}_k$ and standard deviation $\sigma_{r_k}$ were computed using NaN-aware estimators, excluding unvisited bins from the normalization. An active-bin filter was applied, retaining only spatial bins for which at least $\max(2, \lfloor N/3 \rfloor)$ neurons had non-zero firing rates; sessions with fewer than 30 active bins were excluded.

\subsection{The Shesha metric: geometric stability of the neural manifold}
\label{app:shesha}

We quantified the geometric stability of the hippocampal spatial code using the Shesha metric~\citep{raju2026geometric}, a split-half representational dissimilarity matrix (RDM) correlation framework~\citep{Kriegeskorte2008}, with pairwise deletion of unvisited spatial bins.

\paragraph{Feature-split stability ($\text{Shesha}_{\text{FS}}$).} The population of $N$ neurons was randomly partitioned into two equal halves ($N/2$ neurons each). For each half, an RDM was computed over all active spatial bins using \textit{masked cosine distance} with pairwise deletion. Let $V_{ij}^{(h)}$ denote the set of neurons in half~$h$ with valid (finite) $z$-scored rates at both bins $i$ and $j$. The masked cosine distance is:
\begin{equation*}
D_{ij}^{(h)} = 1 - \frac{\displaystyle\sum_{k \in V_{ij}^{(h)}} x_{ik}^{(h)}\, x_{jk}^{(h)}}{\sqrt{\displaystyle\sum_{k \in V_{ij}^{(h)}} \bigl(x_{ik}^{(h)}\bigr)^{\!2}} \;\; \sqrt{\displaystyle\sum_{k \in V_{ij}^{(h)}} \bigl(x_{jk}^{(h)}\bigr)^{\!2}}}
\end{equation*}
where $x_{ik}^{(h)}$ is the $z$-scored response of neuron~$k$ (from half~$h$) at spatial bin~$i$. Pairs with $|V_{ij}^{(h)}| < 2$ were assigned \texttt{NaN} and excluded from downstream correlations. This pairwise deletion prevents unvisited bins (retained as \texttt{NaN}) from inflating similarity through shared zeros. The Shesha score is the Spearman correlation between the valid upper-triangular entries of the two RDMs:
\begin{equation*}
\text{Shesha} = \rho_{\text{Spearman}}\!\bigl(\text{vec}(\mathbf{D}^{(1)}_{\text{valid}}), \, \text{vec}(\mathbf{D}^{(2)}_{\text{valid}})\bigr)
\end{equation*}
computed only over index pairs $(i,j)$ where both $D_{ij}^{(1)}$ and $D_{ij}^{(2)}$ are finite. This was repeated for $n_{\text{splits}} = 100$ random neuron partitions with a fixed random seed (seed $= 320$), and the mean correlation was taken as the final Shesha score. All active bins were used (no subsampling). The Tier~2 E/I pipeline passes these parameters explicitly to \texttt{feature\_split}; the Tier~1 pipeline calls \texttt{feature\_split} with the \texttt{shesha-geometry} package defaults, which are $n_{\text{splits}} = 100$, \texttt{metric} $=$ \texttt{'cosine'}, and \texttt{seed} $= 320$, matching the values documented above.

\paragraph{Sample-split stability ($\text{Shesha}_{\text{SS}}$).} An analogous procedure was applied using sample splits rather than feature splits: subsets comprising 40\% of spatial bins were randomly drawn and RDMs computed from each subsample, yielding a measure of representational robustness across spatial subsamples \\(\texttt{sample\_split(X, n\_splits=30, subsample\_fraction=0.4, metric='cosine', seed=320)}).

\subsection{Mantel test for spatial topology}

To assess whether population-level geometric structure faithfully preserves the physical layout of the environment, we computed the Mantel test~\citep{Mantel1967, Mantel1970}. Physical distances were computed as Euclidean distances between the 2D grid coordinates $(r_i, c_i)$ of each active bin:
\begin{equation*}
d_{\text{phys}}(i,j) = \sqrt{(r_i - r_j)^2 + (c_i - c_j)^2}
\end{equation*}
Neural distances were computed as the root-mean-square distance over mutually valid neurons (pairwise deletion):
\begin{equation*}
d_{\text{neural}}(i,j) = \sqrt{\frac{1}{|V_{ij}|}\sum_{k \in V_{ij}} (\tilde{m}_{k,i} - \tilde{m}_{k,j})^2}
\end{equation*}
where $V_{ij} = \{k : \tilde{m}_{k,i} \text{ and } \tilde{m}_{k,j} \text{ are both finite}\}$ and $\tilde{m}_{k,i}$ is the $z$-scored firing rate of neuron~$k$ at bin~$i$. Pairs with $|V_{ij}| < 2$ were excluded. When all neurons have valid data at both bins, this reduces to $\|\tilde{\mathbf{m}}_i - \tilde{\mathbf{m}}_j\|_2 / \sqrt{N}$. The Tier~1 Mantel analysis uses the masked formulation above (pairwise deletion over $V_{ij}$). In the Tier~2 E/I pipeline, rate maps are zero-filled prior to distance computation (\texttt{nan\_to\_num} replaces all \texttt{NaN} entries with zero before calling \texttt{pdist}), so $V_{ij} = \{1, \ldots, N\}$ for all bin pairs and the expression reduces to standard Euclidean distance divided by $\sqrt{N}$ without pairwise deletion. This zero-filling is appropriate in Tier~2 because the active-bin filter has already removed unvisited bins, and the remaining bins have valid rates for all neurons within each cell-type subpopulation. The observed Mantel statistic was the Spearman rank correlation $r_{\text{obs}}$ between all valid $\binom{n}{2}$ pairwise distances. Statistical significance was assessed against a null distribution of 1{,}000 permutations in which bin labels were randomly shuffled, preserving marginal distributions. Only bins with at least 100 valid distance pairs were analyzed; the reported $p$-value is $p = (\text{number of permutations with}\ r_{\text{perm}} \geq r_{\text{obs}}) / n_{\text{perm}}$.

\subsection{Valiant's Stable Memory Allocator theory and empirical operationalization}
\label{app:valiant-sma}

\paragraph{Theoretical framework.}
A central theoretical motivation for our analysis of hippocampal spatial codes derives from Valiant's proposal that a principal function of the mammalian hippocampus is to serve as a \textit{stable memory allocator} (SMA) for cortex~\citep{Valiant2012}. In this framework, cortex is the main locus of information storage, but hippocampus is required to identify the specific set of cortical neurons at which each new ``chunk''---a conjunctive combination of previously stored items---will be represented. Crucially, the number of neurons allocated to each new chunk must be controlled within a narrow range; otherwise, successive levels of hierarchical memory allocation will cause the allocated populations to either vanish to zero or fill the cortex, rendering the system computationally unstable.

Valiant formalizes this requirement through three properties that a biologically plausible SMA circuit must satisfy simultaneously:

\begin{enumerate}
    \item \textbf{Stability.} For a wide range of input activity levels (densities) spanning more than an order of magnitude---for example, from $\text{Dense}(\mathbf{u}) \in [0.002, 0.025]$---the output density $\text{Dense}(f(\mathbf{u}))$ must fall within a narrow range (e.g., within 1\% of a target value $p^*$). Formally, a circuit has $\epsilon$-stability in the range $[q,s]$ if there exists a fixed $p$ such that for any input $\mathbf{u}$ with density in $[q,s]$, $\text{Dense}(f(\mathbf{u})) \in [p - \epsilon, p + \epsilon]$.

    \item \textbf{Continuity.} If two input patterns $\mathbf{u}$ and $\mathbf{v}$ are similar (Hamming distance $\text{Ham}(\mathbf{u},\mathbf{v})$ small), their outputs $f(\mathbf{u})$ and $f(\mathbf{v})$ should also be similar: $\text{Ham}(f(\mathbf{u}), f(\mathbf{v})) \leq \gamma \cdot \text{Ham}(\mathbf{u}, \mathbf{v})$. This ensures noise tolerance---small perturbations in neural activity do not catastrophically alter the allocation.

    \item \textbf{Orthogonality.} If two inputs differ substantially, their outputs must also differ sufficiently to be treated by cortex as distinct items: $\text{Ham}(f(\mathbf{u}), f(\mathbf{v})) \geq \delta \cdot \text{Ham}(\mathbf{u}, \mathbf{v})$. This prevents catastrophic interference between memories.
\end{enumerate}

\paragraph{Circuit construction.}
Valiant demonstrates that all three properties can be realized by a shallow feedforward network of two or three layers with random connectivity. The key construction uses a threshold function $x + y + z - 2t \geq 1$ on four randomly chosen inputs, where $x$, $y$, $z$ are excitatory and $t$ is inhibitory. If fraction $p$ of inputs are active, the probability that any output neuron fires is:
\begin{equation*}
h(p) = (1-p)\bigl(1-(1-p)^3\bigr) + p^4 = 4p^3 - 6p^2 + 3p
\end{equation*}
This function has a fixed point at $p^* = 1/2$ with derivative $h'(p^*) = 0$, guaranteeing super-exponential convergence. For biologically realistic low-density regimes ($p \approx 0.01$), the construction is extended to $x + y + z - 2t \geq 1$ where $t = \mathbf{1}[(t_1 + \cdots + t_k) \geq 1]$ with $k \approx (\ln 3)/p$ inhibitory inputs, yielding a fixed point near the target density with derivative $h'(p) \to 1 - \ln 3 \approx -0.099$. The construction achieves 3-continuity (output disagreement at most 3 times input disagreement per layer) and $(0.719\ldots)^i$-orthogonality after $i$ layers throughout the input range, with expansion approaching $3/2$ for small perturbations.

\paragraph{Biological correspondences.}
The SMA architecture maps onto the known anatomy of the hippocampal system: the entorhinal cortex serves as the down/up transformer (compressing $\sim$10$^{10}$ cortical neurons to $\sim$10$^7$ hippocampal neurons and back), while the intrinsic hippocampal circuitry (dentate gyrus $\to$ CA3 $\to$ CA1) realizes the inner SMA. The construction requires effectively random connectivity between layers---consistent with the observed decorrelation of place fields among anatomically neighboring hippocampal neurons~\citep{Thompson1990, Redish2001}---and operates at sparse activity levels ($p \approx 0.01$), consistent with recordings from medial temporal lobe showing densities of 0.005--0.012~\citep{Waydo2006}. Neurogenesis in the dentate gyrus can be interpreted within this framework as a mechanism for time-stamping allocations: the steady replacement of input neurons changes the SMA function at a controlled rate, bounded by the continuity property, such that the same stimulus encountered months apart will be allocated to a sufficiently different cortical set to support distinct episodic memories.

\paragraph{Predictions for spatial codes.}
The SMA theory generates three specific, testable predictions for hippocampal spatial representations that we operationalize in this study. First, the \textit{number of neurons allocated to represent each spatial location should be controlled within a narrow range} (stability), predicting low CV in population overlap counts across spatial bins. Second, the \textit{territory assigned to each neuron should be consistent in size} (a corollary of stability applied to the neuron level), predicting low CV in place field areas. Third, the \textit{allocation pattern should be reproducible across independent subsamples of the population} (continuity), predicting high split-half reliability of per-bin neuron counts. In a food-caching species that must reliably encode thousands of cache locations, these SMA properties should be under stronger selective pressure---and hence more pronounced---than in a non-caching species.

\paragraph{Empirical metrics.}
Three metrics were designed to isolate allocation stability from tuning quality, each directly linked to a specific SMA prediction:

\paragraph{Place field detection.} For each neuron, the 2D rate map was smoothed with a Gaussian kernel ($\sigma = 1.0$ bins). Place fields were defined as contiguous regions where the smoothed rate exceeded 30\% of the neuron's peak rate. Connected components were identified via \texttt{scipy.ndimage.label}; fragments smaller than 4 bins were discarded as noise. Neurons with peak rates below 0.5~Hz were classified as non-spatial.

\paragraph{Metric 1: Field-size consistency.} The CV of place field areas (in bins) across all neurons with detected fields was computed. A low CV indicates that each neuron is allocated a territory of consistent size, as predicted by a Valiant-like memory allocation model.

\paragraph{Metric 2: Population overlap consistency.} For each spatial bin falling within at least one neuron's place field, we counted the number of co-active neurons (i.e., neurons whose field mask covered that bin). The CV of this count across field bins quantifies the consistency of ensemble recruitment; low CV indicates that each location activates a similarly sized neural ensemble.

\paragraph{Metric 3: Split-half allocation reliability.} The neuron population was randomly partitioned 50 times. For each split, we counted the number of field-active neurons per bin in each half and computed the Spearman correlation between the two halves. The mean correlation across splits measures the reproducibility of the spatial allocation pattern. The one-tailed Mann--Whitney test for this metric used \texttt{alternative='less'} (chickadee $<$ finch), because the double-dissociation predicts that caching birds exhibit \textit{lower} allocation reliability---their caching neurons actively shift ensemble membership across trials, consistent with a geometric attractor mechanism rather than fixed ensemble assignment.

\subsection{Alternative geometric stability benchmarks}

To address concerns about whether the Shesha metric, as a novel measure, might yield results not reproducible by established alternatives, we implemented two additional stability metrics and verified that all three agree directionally on the same sessions.

\paragraph{Test-retest population vector (PV) correlation.} The neuron population was randomly partitioned into two halves ($n_{\text{splits}} = 100$, seed $= 320$). For each split, the population vector---the mean firing rate per spatial bin, averaged over all neurons in that half---was computed for each half. The Spearman rank correlation between the two PV profiles was computed over bins where at least one half had non-zero activity ($\geq 30$ active bins required). The mean across splits is the PV stability score. This metric is the simplest interpretable alternative: it measures whether the aggregate spatial preference of a random neuron subset is reproducible, without making any geometric assumptions.

\paragraph{Canonical correlation analysis (CCA) stability.} Neuron halves were used directly as feature matrices. For each split, let $\mathbf{X}_1 \in \mathbb{R}^{n_{\text{bins}} \times k_1}$ and $\mathbf{X}_2 \in \mathbb{R}^{n_{\text{bins}} \times k_2}$ denote the $z$-scored activity matrices of the two halves over active bins, where $k_1, k_2$ are the respective half-sizes. Regularized within-group covariance matrices were formed:
\begin{equation*}
\mathbf{C}_{11} = \frac{\mathbf{X}_1^\top \mathbf{X}_1}{n_{\text{bins}}} + 10^{-6}\mathbf{I}, \quad \mathbf{C}_{22} = \frac{\mathbf{X}_2^\top \mathbf{X}_2}{n_{\text{bins}}} + 10^{-6}\mathbf{I}, \quad \mathbf{C}_{12} = \frac{\mathbf{X}_1^\top \mathbf{X}_2}{n_{\text{bins}}}
\end{equation*}
The cross-covariance was whitened using Cholesky factors of the inverse within-group covariances:
\begin{equation*}
\mathbf{T} = \mathbf{C}_{11}^{-1/2}\,\mathbf{C}_{12}\,\mathbf{C}_{22}^{-1/2}
\end{equation*}
The singular values $\sigma_1 \geq \sigma_2 \geq \cdots$ of $\mathbf{T}$, clipped to $[0, 1]$, are the canonical correlations. The CCA stability score is the mean of the top $d = \min(3, k_1 - 1, k_2 - 1)$ canonical correlations, averaged over $n_{\text{splits}} = 50$ random neuron partitions. The number of components was chosen adaptively---rather than requiring a fixed minimum half-size, $d$ was reduced whenever halves were small---to avoid discarding all splits for low-neuron sessions (which would yield spurious NaN rather than a meaningful low score). A Cholesky decomposition failure (near-singular covariance, caught as \texttt{LinAlgError}) discarded only that split.

\paragraph{Procrustes-aligned split-half error.} For each random neuron split, the pairwise cosine distance matrix between \textit{spatial bins} (not neurons) was computed from each half. Let $\mathbf{D}^{(h)} \in \mathbb{R}^{n_{\text{bins}} \times n_{\text{bins}}}$ be the squareform cosine distance matrix from half $h$, where distances are taken between rows of $\mathbf{X}_h^\top$ (i.e., between bin-level activity patterns projected into the half-neuron space). An orthogonal Procrustes alignment was applied to $\mathbf{D}^{(2)}$ relative to $\mathbf{D}^{(1)}$, finding the rotation $\mathbf{R}^* = \arg\min_{\mathbf{R}^\top\mathbf{R}=\mathbf{I}} \|\mathbf{D}^{(1)} - \mathbf{D}^{(2)}\mathbf{R}\|_F$. The normalized Frobenius residual after alignment is:
\begin{equation*}
\varepsilon = \frac{\|\mathbf{D}^{(1)} - \mathbf{D}^{(2)}\mathbf{R}^*\|_F}{\|\mathbf{D}^{(1)}\|_F}
\end{equation*}
Procrustes stability is reported as $1 - \bar{\varepsilon}$ (mean residual over $n_{\text{splits}} = 100$ splits), so that higher values indicate better geometric reproducibility. This metric aligns the full pairwise distance structure directly---without dimensionality reduction or embedding---and is therefore embedding-free.

\paragraph{Concordance check.} For each of the three metrics (Shesha, PV correlation, CCA stability), a one-sided Mann--Whitney $U$ test (chickadee $>$ finch) was computed over the session-level scores. Directional consistency (chickadee mean $>$ finch mean) was assessed for each metric independently and recorded in a concordance summary table. All three metrics agreeing directionally constitutes the primary concordance result. Procrustes-aligned split-half stability was additionally computed as a supplementary benchmark (see above) but was not included in the primary concordance set.

\subsection{Within-session representational drift}

Temporal stability of individual neuron rate maps was assessed using the cross-correlation of first-half versus second-half maps (\texttt{xcorr\_map}), as provided in the original dataset. Species differences were tested using one-sided Mann--Whitney $U$ tests (chickadee $>$ finch).

\subsection{Anterior--posterior gradient}

For chickadee sessions with histologically confirmed recording-site subdivisions (available in the \texttt{subdivision} column of the dataset), Shesha scores were computed per subdivision and compared using Mann--Whitney $U$ tests to assess whether geometric stability varies along the anterior--posterior axis of the hippocampal formation.

\subsection{Neuron-count-matched downsampling control}

To control for the possibility that species differences in Shesha reflect differences in recorded neuron counts rather than genuine geometric stability, we performed a neuron-count-matched downsampling analysis. The target neuron count $N_{\text{target}}$ was set to the median number of neurons per session in the zebra finch dataset. For each chickadee session with at least $N_{\text{target}}$ neurons, a random subset of $N_{\text{target}}$ neurons was drawn without replacement, Shesha was recomputed on the subsampled population, and this procedure was repeated 50 times per session. The per-session mean across subsamples was taken as the neuron-matched Shesha score. The distribution of matched chickadee scores was compared to the original finch distribution using one-sided Mann--Whitney $U$ tests (chickadee $>$ finch) and bootstrapped Cohen's $d$ with 95\% confidence intervals ($n_{\text{boot}} = 10{,}000$). Results are reported alongside the unmatched analysis in the negative controls output.

\subsection{Tier~1 negative controls for geometric stability}

To rule out the possibility that the elevated Shesha scores in chickadees reflect single-neuron tuning properties rather than genuine population-level geometric structure, two additional negative controls were applied to every session passing the minimum neuron criterion ($N \geq 5$):

\paragraph{Circular shift control.} Each neuron's one-dimensional rate map was circularly shifted by a uniformly random offset drawn independently for each neuron. This preserves the marginal firing-rate distribution and spatial smoothness of individual neurons but destroys the inter-neuron spatial relationships that define population geometry. For each session, 20 shifted populations were generated and Shesha was computed on each; the per-session control score was the mean across the 20 iterations.

\paragraph{Map shuffle control.} Each neuron's rate-map bins were randomly permuted (without replacement), independently across neurons. This eliminates spatial continuity within each neuron's map and therefore destroys manifold smoothness at both the single-neuron and population levels. As above, 20 shuffled populations were generated per session and the per-session control score was the mean across iterations.

If the species difference in Shesha reflects population geometry, original scores should substantially exceed both controls. A collapse of the species difference under either control would indicate that single-neuron statistics, rather than coordinated population structure, drive the observed effect.

\subsection{Excitatory-inhibitory circuit analysis}
\label{app:ei-circuit}

\paragraph{E/I decomposition.} For each chickadee session meeting the minimum neuron criterion ($\geq 3$ E neurons and $\geq 3$ I neurons), population matrices were constructed separately for excitatory ($\mathbf{M}_E$), inhibitory ($\mathbf{M}_I$), and combined ($\mathbf{M}_{\text{all}}$) subpopulations. Shesha was computed independently for each.

\paragraph{Residual contribution analysis.} To isolate the unique geometric information carried by each cell type, we subtracted the mean spatial map of the opposite type from each neuron's rate map prior to computing Shesha. The shared variance between E and I populations was quantified as the Spearman correlation between their mean spatial maps over active bins.

\paragraph{Partial information decomposition (PID).} A simplified PID was computed using Shesha as a proxy for geometric information content. Redundant information was defined as $\min(\text{Shesha}_E, \text{Shesha}_I)$; unique E and I contributions as the respective excesses over the redundant component; and synergistic information as $\max(0, \text{Shesha}_{\text{all}} - (\text{Shesha}_E + \text{Shesha}_I - \text{redundant}))$, capturing geometric structure that emerges only from the combined population.

\paragraph{E-I coordination index.} The Spearman rank correlation between per-session $\text{Shesha}_E$ and $\text{Shesha}_I$ values was computed to assess whether E and I geometric stability co-vary across recording sessions.

\paragraph{Spatial frequency analysis.} Mean population rate maps were computed for E and I populations in each session. A 2D fast Fourier transform was applied and the power spectrum was radially averaged. The Spearman correlation between the normalized E and I power spectra quantified spectral similarity. Dominant spatial frequencies were identified as the non-DC peak of the radial power spectrum.

\paragraph{Dimensionality analysis.} The intrinsic dimensionality of each subpopulation was estimated via PCA on the transposed (bins $\times$ neurons) rate map matrix, recording the number of principal components required to explain 95\% of the variance. Subspace overlap between E and I representations was quantified as the absolute Pearson correlation between the first principal components of each subpopulation.

\paragraph{Temporal dynamics.} To distinguish between the \textit{geometric} stability captured by the feature-split Shesha metric (Shesha$_\text{FS}$) and a complementary \textit{temporal} stability reflecting trial-to-trial consistency, we computed a sample-split variant of Shesha (Shesha$_\text{SS}$) for each E and I subpopulation. Sample-split stability was computed by randomly subsampling 40\% of the spatial bins across 30 bootstrap iterations ($\texttt{subsample\_fraction} = 0.4$, $n_\text{splits} = 30$, cosine metric, seed 320) and correlating the pair of resulting RDMs, yielding a measure of how stable the representational geometry is across different subsets of the spatial context rather than across different subsets of neurons. For each session, Shesha$_\text{FS}$ (feature-split, neurons partitioned) and Shesha$_\text{SS}$ (sample-split, bins subsampled) were computed separately for the E and I populations, and the ratio Shesha$_\text{FS}$ / Shesha$_\text{SS}$ was reported as a scale-free index of how much the geometric structure depends on the spatial coverage relative to the neuron identity. Per-session values for both stability types and the ratio are exported to supplementary data. This analysis is restricted to chickadee sessions (the finch inhibitory population is insufficient for separate E/I comparisons) and uses the global minimum neuron threshold ($n_E, n_I \geq 3$).

\paragraph{E/I principal subspace angle analysis.} To provide a better-powered and more direct test of the claim that E and I populations occupy distinct representational subspaces---complementing the underpowered session-level E/I Shesha correlation---we computed the principal angles between the E and I principal subspaces. For each session, the top-$d$ left singular vectors of $\mathbf{M}_E^\top$ and $\mathbf{M}_I^\top$ were extracted via SVD (where rows are spatial bins and columns are neurons), yielding orthonormal bases $\mathbf{Q}_E, \mathbf{Q}_I \in \mathbb{R}^{n_{\text{bins}} \times d}$ representing the dominant directions of spatial activity variation in each population. The principal angles $\theta_1 \leq \theta_2 \leq \cdots \leq \theta_d$ between the two subspaces were computed from the singular values $\sigma_1 \geq \cdots \geq \sigma_d$ of $\mathbf{Q}_E^\top \mathbf{Q}_I$:
\begin{equation*}
\theta_i = \arccos(\sigma_i), \quad \sigma_i = \text{clip}(\sigma_i, -1, 1)
\end{equation*}
with $d = \min(3, n_E - 1, n_I - 1, n_{\text{bins}} - 1)$. Sessions with $\min(n_E, n_{\text{bins}}) < d$ were excluded. Mean principal angle across the top-$d$ dimensions was computed per session; angles approaching 90$^\circ$ indicate orthogonal (non-overlapping) subspaces, while small angles indicate shared representational structure. This analysis does not depend on single-session correlations and is substantially better powered than the session-level E/I Shesha coordination correlation ($r = -0.333$, $p = 0.381$), providing an independent test of whether the claim of orthogonal E/I dynamics is supported by the data. The minimum neuron requirement for this analysis is $n_E \geq 3$ and $n_I \geq 3$ (matching the global session-inclusion criterion \texttt{MIN\_NEURONS\_TYPE}$\,= 3$), because SVD is well-defined for any matrix with at least as many columns as the target subspace dimension. This yields $n = 21$ sessions and is deliberately the least restrictive threshold used in Tier~2. As a sensitivity check, the analysis was repeated with a strict threshold of $n_E \geq 4$ and $n_I \geq 4$ (matching the Shesha split-half requirement), reducing the sample to the same $n = 12$ sessions used by the paired synergy test; principal angle estimates from both thresholds are reported in the supplementary data and should agree if the result is robust to the three borderline sessions.

\paragraph{Paired full-population vs.\ E-only synergy test.} The claim that the combined E/I population exhibits higher geometric stability than the excitatory population alone---i.e.\ that inhibitory neurons contribute synergistically to manifold structure rather than merely adding noise---requires a direct paired within-session test. This analysis operates on a subset of the sessions used for subspace angles ($n = 12$ vs.\ $n = 21$) for the following reason: the Shesha split-half procedure requires at least 4 neurons per subpopulation to form two non-trivial halves of $\geq 2$ neurons each, so sessions with exactly $n_E = 3$ pass the global inclusion criterion (\texttt{MIN\_NEURONS\_TYPE}$\,= 3$) and therefore contribute to the subspace angle analysis, but return NaN from the split-half computation and are consequently excluded from the paired test. The nine-session difference reflects this deliberate asymmetry: SVD-based subspace estimation is valid at $n_E = 3$ (one vector per subspace dimension), while a split-half RDM correlation is degenerate with only one neuron per half. All sessions satisfying each analysis's respective minimum neuron criterion were included; no sessions were selectively excluded.

As a cross-check on both analyses simultaneously, we ran Shesha with a relaxed threshold ($n_E \geq 3$, allowing sessions with a degenerate one-neuron half) and the subspace angle analysis with a strict threshold ($n_E \geq 4$). This produces two additional sample-matched comparisons: (i) a relaxed Shesha paired synergy test at $n = 21$ sessions, matching the primary subspace angle sample; and (ii) a strict subspace angle analysis at $n = 12$ sessions, matching the primary paired synergy test. Concordance across all four variants---primary strict Shesha, relaxed Shesha, primary relaxed subspace angles, strict subspace angles---constitutes the robustness check for the threshold-sensitivity concern.

For each session in which both $\text{Shesha}_{\text{full}}$ and $\text{Shesha}_{E}$ were computable (non-NaN), we formed the signed difference $\Delta_s = \text{Shesha}_{\text{full},s} - \text{Shesha}_{E,s}$. The number of sessions with $\Delta_s > 0$ was recorded, and the one-sided Wilcoxon signed-rank test ($H_1{:}\ \text{Shesha}_{\text{full}} > \text{Shesha}_{E}$) was applied to the paired observations. The Wilcoxon test was chosen over a paired $t$-test because Shesha scores are bounded in $[-1, 1]$ and their differences cannot be assumed normally distributed at the small session counts available. This test is strictly more informative than the unpaired E vs.\ I comparison ($p = 0.939$), which addresses a different contrast; the synergy claim requires the paired Full vs.\ E-only test. Per-session values of $\text{Shesha}_{\text{full}}$, $\text{Shesha}_{E}$, $\text{Shesha}_{I}$, and $\Delta_s$ are reported in the supplementary data alongside the test statistic $W$ and one-sided $p$-value.

\paragraph{Negative controls.}
\begin{enumerate}
    \item \textit{Anti-correlated I population:} An artificial I population was generated by sign-inverting the E population and rescaling to non-negative values, establishing an upper bound on E/I conflict.
    \item \textit{Random I pairing:} Each session's E neurons were paired with I neurons drawn from a different session, breaking within-session E-I coordination while preserving marginal statistics.
    \item \textit{Scaled noise injection:} Graded Gaussian noise ($\sigma \in \{0.0, 0.1, 0.2, 0.5, 1.0\}$) was added to the I population's rate maps to identify the noise level at which geometric stability degrades.
\end{enumerate}

\paragraph{Statistical inference.}
Species comparisons employed Mann--Whitney $U$ tests (two-sided or one-sided as specified). The paired Full vs.\ E-only synergy test used the one-sided Wilcoxon signed-rank test on session-level differences ($H_1{:}\ \Delta > 0$). Within-chickadee E/I differences were further assessed using:
\begin{itemize}
    \item Simple bootstrap confidence intervals ($n_{\text{boot}} = 10{,}000$; 95\% percentile CIs), appropriate for within-species comparisons.
    \item Cohen's $d$ effect sizes with bootstrap CIs ($n_{\text{boot}} = 10{,}000$).
    \item Non-parametric permutation tests ($n_{\text{perm}} = 10{,}000$).
    \item Jackknife resampling to verify that results were not driven by single influential sessions.
    \item Metropolis--Hastings Markov chain Monte Carlo (MCMC; $n_{\text{MCMC}} = 1{,}000$ samples, 50\% burn-in, yielding 500 post-burn-in samples) for Bayesian credible intervals on the E-I difference, assuming a Gaussian likelihood with uniform priors.
\end{itemize}

\subsection{Computational model of topological capacity}
\label{app:computational-model}

\paragraph{Population code generation.} Synthetic neural populations were generated as follows. $N = 500$ neurons were assigned uniformly random preferred locations on a 1D circular track $[0, 1)$ (the supplementary phase-transition boundary and extended temporal-drift simulations used $N = 400$ for computational tractability; all other parameters were identical). Each neuron's tuning curve was modeled as a Gaussian bump:
\begin{equation*}
r_j(x) = \exp\!\Bigl(-\frac{d(x, x_j^*)^2}{2\sigma^2}\Bigr), \quad d(x, y) = \min(|x - y|, 1 - |x - y|)
\end{equation*}
where $x_j^*$ is the preferred location and $\sigma = \rho/2$ with sparsity $\rho = 0.15$ (standard configuration). Functional noise was added as i.i.d.\ Gaussian perturbations ($\sigma_{\text{noise}} = 0.2$), and negative rates were rectified. The \textit{topology strength} parameter $\tau \in [0, 1]$ controlled the fraction of neurons retaining their topological tuning curves; the remaining $(1 - \tau)N$ neurons had their tuning curves randomly permuted across locations. Population vectors were L2-normalized. Three canonical regimes were defined:

\begin{center}
\begin{tabular}{lll}
\textbf{Regime} & \textbf{Topology} $\tau$ & \textbf{Biological analogue} \\
\hline
Crystal & 1.0 & Chickadee (food-cacher) \\
Mist & 0.5 & Zebra finch (non-cacher) \\
Noise & 0.0 & Null (no topology) \\
\end{tabular}
\end{center}

The achieved topology fidelity of each regime was verified by computing the Mantel~$r$ (Spearman correlation between pairwise physical and neural Euclidean distances) at $M = 100$ locations, averaged over 15 independent random seeds (320, 1991, 9, 7258, 7, 2222, 724, 3, 12, 108, 18, 11, 1754, 411, 103) to obtain stable estimates.

\paragraph{Decoders.} Four decoding strategies were compared:
\begin{enumerate}
    \item \textit{Noisy nearest-neighbor (NN):} Independent Gaussian readout noise ($\sigma_{\text{read}} = 0.3$) was added to each element of the population vector. The noisy vector was then matched against the noiseless template library by Euclidean distance (self-match excluded via $D_{ii} = \infty$). Decoding performance was quantified as the \textit{misclassification rate}---the fraction of locations for which the nearest-neighbor template did not correspond to the true location---rather than a continuous circular distance. This discrete error metric is unambiguous when the stimulus set is a finite set of stored locations and avoids the implicit assumption of a smoothly varying error surface.
    \item \textit{Linear (Ridge regression):} A RidgeCV decoder ($\alpha \in \{0.1, 1.0, 10.0\}$) was trained on noiseless population vectors and evaluated on noisy inputs.
    \item \textit{Support vector regression (SVR):} An $\epsilon$-SVR with RBF kernel ($C = 1.0$, $\epsilon = 0.1$) was trained on noiseless codes and tested on noisy readout.
    \item \textit{Ideal Bayesian:} Maximum a posteriori estimation under a Gaussian noise model with uniform prior, $\hat{x} = \arg\max_x P(\mathbf{r}|x) = \arg\min_x \sum_j (r_j - \mu_j(x))^2 / (2\sigma_{\text{read}}^2)$.
\end{enumerate}
For the NN decoder, error was the misclassification rate $\epsilon_{\text{NN}} = 1 - \frac{1}{M}\sum_{i=1}^{M}\mathbf{1}[\hat{x}_i = x_i]$, where $\hat{x}_i$ is the nearest-neighbor template and $x_i$ is the true location. For the regression-based decoders (Ridge, SVR, Bayesian), error was the mean circular distance $\epsilon = \min(|x - \hat{x}|,\, 1 - |x - \hat{x}|)$ averaged over test locations.

\paragraph{Memory capacity.} Capacity was tested by increasing the number of stored locations $M \in \{10, 20, 30, \ldots, 1000\}$ and measuring the NN misclassification rate at each memory load. The critical capacity was defined as the memory load at which the misclassification rate exceeded a threshold set at the midpoint between crystal and noise error rates at $M = 100$ locations; interpolation between adjacent loads refined the estimate.

\paragraph{Negative controls.}
\begin{enumerate}
    \item \textit{Matched-random control:} Random codes were generated by searching over topology strengths to match the Mantel $r$ of the target code ($|\Delta r| < 0.05$), controlling for total distance-preserving structure.
    \item \textit{Anti-topological control:} The distance--similarity relationship was inverted ($d_{\text{inv}} = 0.5 - d$), producing codes where spatially distant locations have similar representations.
    \item \textit{Dimension-matched control:} Random Gaussian codes were projected onto a subspace of specified dimensionality via SVD truncation, then sparsified and normalized, controlling for effective rank.
\end{enumerate}

\paragraph{Noise robustness.}
The topology advantage was stress-tested under three noise models:
\begin{itemize}
    \item \textit{Additive Gaussian noise:} Readout noise scaled from $\sigma = 0.1$ to $1.0$ (10 levels).
    \item \textit{Multiplicative (Poisson-like) noise:} Signal-proportional noise, $\tilde{r} = r \cdot (1 + \alpha \cdot z)$, $z \sim \mathcal{N}(0,1)$, with gain factor $\alpha \in [0.1, 2.0]$ (10 levels).
    \item \textit{Correlated noise:} A fraction $c \in \{0.0, 0.2, 0.4, 0.6, 0.8\}$ of the noise variance was shared across neurons: $\boldsymbol{\eta} = \sqrt{c}\,\eta_{\text{common}} + \sqrt{1-c}\,\boldsymbol{\eta}_{\text{indep}}$.
\end{itemize}
The \textit{critical noise threshold}---the readout noise level at which the crystal--mist performance difference fell below a threshold of 0.1---was identified by sweeping 20 noise levels from 0.1 to 2.0.

\paragraph{Information-theoretic bounds.}
A theoretical lower bound on capacity was computed using Shannon channel capacity under a Gaussian channel model:
\begin{equation*}
C_{\text{lower}} = \frac{1}{2} \log_2\!\left(1 + \frac{\rho N}{\sigma_{\text{read}}^2}\right) \quad \text{bits}
\end{equation*}
where $\rho N$ approximates the sparse signal power. An upper bound was estimated as $C_{\text{upper}} = \log_2(N/\rho)$, based on the maximum number of distinguishable patterns. Empirical information content was estimated from nearest-neighbor distances in the neural space.

\paragraph{Topology sweep.} A parametric sweep over topology strength ($\tau \in \{0.0, 0.1, 0.2, \ldots, 1.0\}$, 11 levels) was conducted at a fixed memory load of 100 locations with 250 independent trials per level. Means and 95\% percentile confidence intervals were computed for both decoding error and Mantel~$r$.

\paragraph{2D open-field arena validation.} The primary capacity simulation used a 1D circular track to keep the geometry tractable across large parameter sweeps. To verify that the key claims generalize to the open-field arena geometry used in the empirical recordings, we ran a targeted 2D simulation with 210 network configurations (10 trials $\times$ 7 memory loads $\times$ 3 regimes). Tuning curves were 2D isotropic Gaussian place fields:
\begin{equation*}
r_j(\mathbf{x}) = \exp\!\left(-\frac{\|\mathbf{x} - \mathbf{c}_j\|^2}{2\sigma^2}\right), \quad \mathbf{x}, \mathbf{c}_j \in [0, 1]^2
\end{equation*}
where $\mathbf{c}_j \sim \text{Uniform}([0,1]^2)$ is the preferred location of neuron $j$ and $\sigma = \rho/2$ with $\rho = 0.15$ (matching the 1D configuration and the normalized empirical 40$\times$40 arena). Physical distances were 2D Euclidean rather than circular. The noisy nearest-neighbor error was the misclassification rate (fraction of 2D locations decoded to the wrong template), consistent with the updated error metric used throughout Tier~3. All other simulation parameters ($N = 500$, $\sigma_{\text{noise}} = 0.2$, $\sigma_{\text{read}} = 0.3$) were held identical to the 1D case to isolate the effect of arena geometry. The Mantel statistic used 2D Euclidean physical distances. We report: (i) the directional ordering of misclassification rates (crystal $\leq$ mist $\leq$ noise) across all seven memory loads; (ii) the critical capacity for each regime; and (iii) a phase diagram of error as a function of both topology strength and memory load, visualizing where the phase transition occurs in 2D parameter space. All results and per-load error rates are exported to supplementary data.

\paragraph{Empirical redundancy.} Redundancy was quantified from the real Payne data as
\begin{equation*}
R = \frac{\sum_{k:\,I_k > 0} I_k}{I_{\text{pop}}}
\end{equation*}
where $I_k$ is the spatial information of individual neuron $k$ and $I_{\text{pop}}$ is the population-level spatial information estimated via cross-validated Ridge regression ($\alpha \in \{0.01, 0.1, 1.0, 10.0\}$; 5-fold CV). The decoded variables were the row and column indices of each spatial bin, and population information was computed as $I_{\text{pop}} = -\frac{1}{2}\log_2(1 - R^2)$ where $R^2$ is the mean cross-validated coefficient of determination across both spatial dimensions. Species differences in redundancy were tested using two-sided Mann--Whitney $U$ tests. Because the redundancy ratio diverges as $I_{\text{pop}} \to 0$, a strict floor of $I_{\text{pop}} \geq 0.05$~bits was applied: sessions below this threshold were excluded from a filtered redundancy analysis, and both unfiltered and filtered results are reported to demonstrate robustness.

\paragraph{Absolute population information with bootstrap confidence intervals.} To supplement the redundancy ratio with interpretable absolute quantities, we additionally report $I_{\text{pop}}$ in bits per session with 95\% bootstrap confidence intervals. For each session, neurons were resampled with replacement ($n_{\text{boot}} = 500$ iterations), $I_{\text{pop}}$ was recomputed on each bootstrap replicate via the same cross-validated Ridge regression procedure, and the 2.5th and 97.5th percentiles of the bootstrap distribution were taken as the CI bounds. Sessions for which the point estimate $I_{\text{pop}}$ was NaN (insufficient active bins for cross-validation) were excluded. Per-session absolute $I_{\text{pop}}$ values are reported alongside the redundancy ratios, enabling readers to verify that the observed redundancy differences are not artifacts of sessions with near-zero $I_{\text{pop}}$.

\subsection{10,000-configuration parameter sweep}

To assess the generality of the topology advantage across biologically plausible parameter regimes, a comprehensive sweep was conducted over $20 \times 25 \times 20 = 10{,}000$ combinations of:
\begin{itemize}
    \item Population size $N \in \{25, 50, 75, \ldots, 500\}$ (20 values)
    \item Sparsity $\rho \in \{0.01, 0.02, 0.03, \ldots, 0.25\}$ (25 values)
    \item Number of independent trials $T \in \{25, 50, 75, \ldots, 500\}$ (20 values)
\end{itemize}
For each combination, all four decoders were evaluated at a fixed memory load of 100 locations under crystal ($\tau = 1.0$), mist ($\tau = 0.5$), and noise ($\tau = 0.0$) regimes. Capacity advantage was defined as the difference in NN error between the random and topological codes. Note that the NN error metric used throughout the parameter sweep is the mean circular distance (not the misclassification rate used in the targeted capacity analysis); this sweep predates the metric update, but the qualitative conclusion---crystal $<$ mist $<$ noise ordering is robust across all 10,000 parameter combinations---is unchanged because both metrics rank the three regimes identically. The sweep was executed in batches of 3{,}500 combinations (to accommodate the Google Colab 24-hour runtime limit) with incremental checkpointing every 100 combinations. Batch results were combined post hoc into a single dataset for analysis. Summary statistics were computed by marginalizing over each parameter dimension.

\subsection{Statistical framework}

All species comparisons (Tier~1) employed one-sided Mann--Whitney $U$ tests (chickadee $>$ finch) where the theoretical prediction specified a direction, and two-sided tests otherwise. An important exception is the split-half allocation reliability metric (Metric 3), for which the one-tailed test direction is chickadee $<$ finch: the double-dissociation hypothesis predicts that caching birds show \textit{lower} allocation reliability (i.e., more dynamic ensemble membership) alongside higher geometric stability. Reversing this direction would test the wrong tail and yield the complement of the correct $p$-value. Within-species E/I comparisons (Tier~2) used simple (non-hierarchical) bootstrap, which is statistically appropriate because no cross-species inference is required. Confidence intervals are reported at the 95\% level throughout. All simulations used fixed random seeds (320) for exact reproducibility. Figures present individual data points alongside summary statistics.

The three alternative stability benchmarks (Shesha, PV correlation, CCA stability; Tier~1) were treated as a pre-specified concordance set: each metric was tested independently, and the primary claim of species difference is supported only if all three agree directionally. This criterion mitigates the risk of over-interpreting any single novel metric; the concordance summary reporting means, $p$-values, and directional consistency flags for each metric is provided in the supplementary data.

\subsection{Software and code availability}

All analyses were implemented in Python~3 using \texttt{NumPy}, \texttt{SciPy}, \texttt{pandas}, \texttt{scikit-learn}, \texttt{matplotlib}, and \texttt{tqdm}. The Shesha metric is implemented in the open-source \texttt{shesha-geometry} package~\citep{shesha2026}, available on the Python Package Index (\url{https://pypi.org/project/shesha-geometry/}). Custom analysis code, including all tiered analyses and the parameter sweep framework, is available at \url{https://github.com/prashantcraju/hippocampal-stability}.

The primary Tier~1 analysis encompasses the Shesha metric, the revised Valiant allocation metrics, the three alternative stability benchmarks (test-retest PV correlation, CCA stability, Procrustes-aligned split-half), the negative controls (circular shift, map shuffle, and neuron-count-matched downsampling), and the concordance verification across all three benchmarks. The Tier~2 E/I analysis is implemented in two complementary pipelines: the primary pipeline (\texttt{tier2\_ei\_stability.py}) produces Shesha by cell type, the Mantel test, spatial information, within-session stability, E/I spatial correlation, principal subspace angle analysis, the paired full-population vs.\ E-only synergy test, and the associated negative controls; a supplementary pipeline (\texttt{tier2\_enhanced.py}) contributes the residual contribution analysis, partial information decomposition, spatial frequency analysis, dimensionality analysis, temporal dynamics (Shesha$_\text{SS}$), and Bayesian MCMC inference. The Tier~3 analysis covers the 1D computational capacity simulation, the 10{,}000-configuration parameter sweep, the 2D open-field arena validation, the topology sensitivity sweep, and the empirical redundancy analysis including per-session absolute $I_{\text{pop}}$ with bootstrap confidence intervals.

\section{Figures}

\begin{figure}[H]
    \centering
    \includegraphics[width=\linewidth]{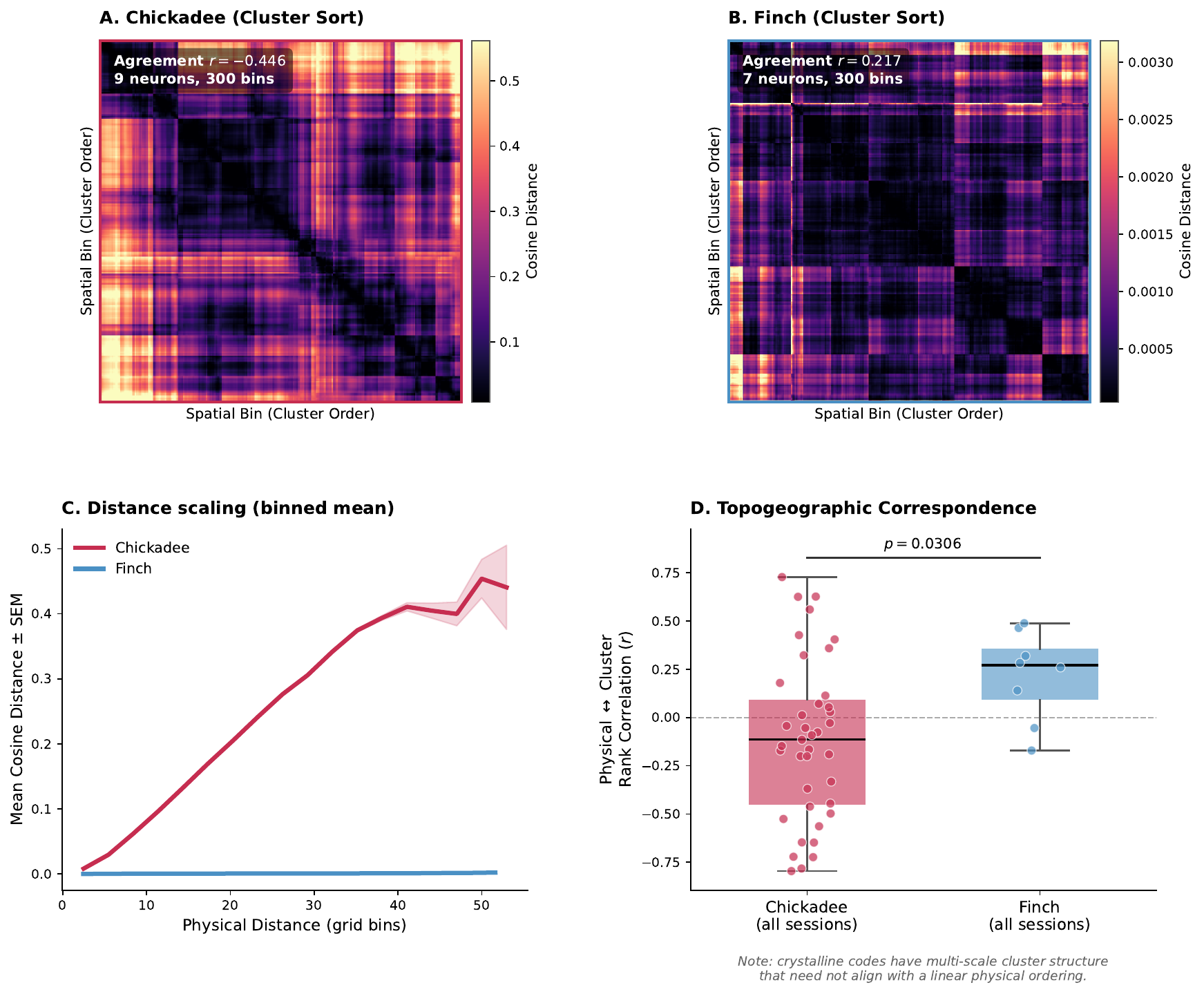}
    \caption{\textbf{Cluster-sorted RDMs and
    topogeographic correspondence.}
    \textbf{(A, B)} RDMs for the best-session chickadee (A) and finch (B), re-sorted by Ward hierarchical clustering on the RDM itself rather than by physical arena order. Physical$\leftrightarrow$cluster agreement
    (Spearman $r$ between physical rank ordering and cluster leaf ordering) is $r = -0.446$ for the chickadee and $r = 0.217$ for the finch.
    \textbf{(C)} Binned mean $\pm$ SEM of neural cosine distance as a function of physical distance between bin pairs (best session per species). The chickadee shows a monotonically increasing physical--neural distance relationship; the finch curve remains near zero across all physical distances, indicating that neural population vectors carry negligible spatial distance information.
    \textbf{(D)} Physical$\leftrightarrow$cluster rank correlation across all sessions (Mann--Whitney $U$, $p = 0.031$). Chickadee sessions exhibit broadly distributed agreement values with a median near zero and a negative tail, while finch sessions show consistently positive agreement. This pattern reflects a qualitative difference in representational geometry: the chickadee hippocampus organizes spatial representations into a rich, multi-scale cluster hierarchy whose dominant axes are not aligned with the cardinal row-major physical ordering used here, whereas the finch cluster structure is weaker overall and more consistent with a simple topographic layout. The chickadee's negative physical$\leftrightarrow$cluster $r$ values thus reflect the complexity of its geometric organization rather than an absence of spatial structure; the high Mantel $r$ values in Figure~\ref{fig:rdm_main}C confirm that topogeographic correspondence is in fact stronger in chickadees at the population level.
    }
    \label{fig:rdm_supp}
\end{figure}

\newpage

\begin{figure}[H]
    \centering
    \includegraphics[width=\linewidth]{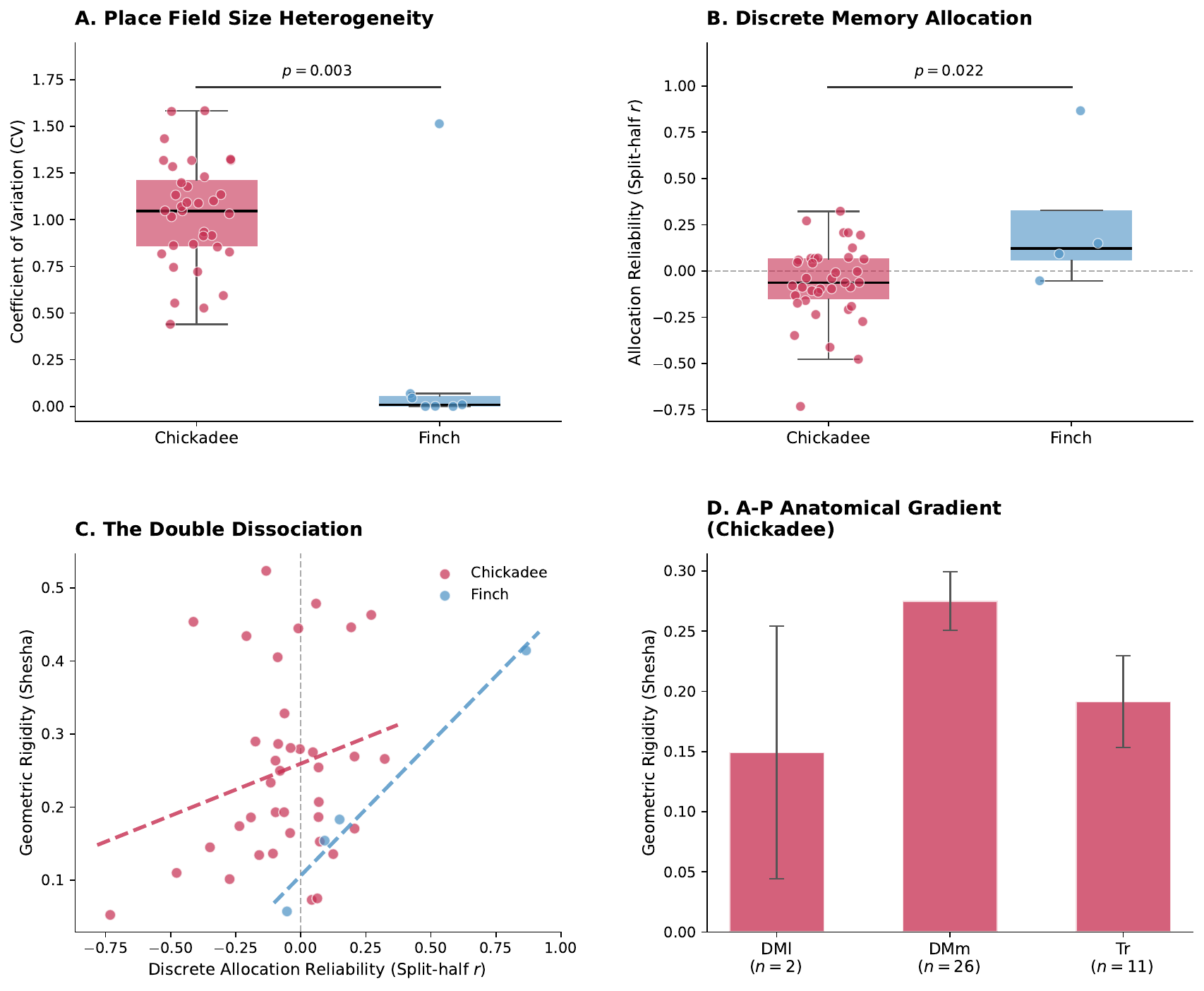}
    \caption{ \textbf{Double dissociation between geometric stability and discrete memory allocation.} 
    \textbf{(A)} Place field size heterogeneity. Caching networks exhibit a significantly higher coefficient of variation (CV) in place field sizes, defying the Stable Memory Allocator (SMA) prediction that individual locations recruit consistent, uniform computational territories. 
    \textbf{(B)} Split-half allocation reliability. The caching hippocampus exhibits near-zero reliability in recruiting the same number of active neurons across subsets, actively refuting the hypothesis of rigid, discrete cellular allocation. 
    \textbf{(C)} The double dissociation. Caching networks achieve extreme geometric rigidity (high Shesha) despite functionally zero discrete allocation reliability. This demonstrates that spatial memory capacity is stored in the continuous topology of the population manifold rather than in rigid, discrete cellular assignments. 
    \textbf{(D)} Anterior-Posterior anatomical gradient of geometric stability within the caching chickadee hippocampus, demonstrating the distribution of manifold rigidity across subdivisions.}
    \label{fig:supp_valiant}
\end{figure}

\newpage

\begin{figure}[H]
    \centering
    \includegraphics[width=\linewidth]{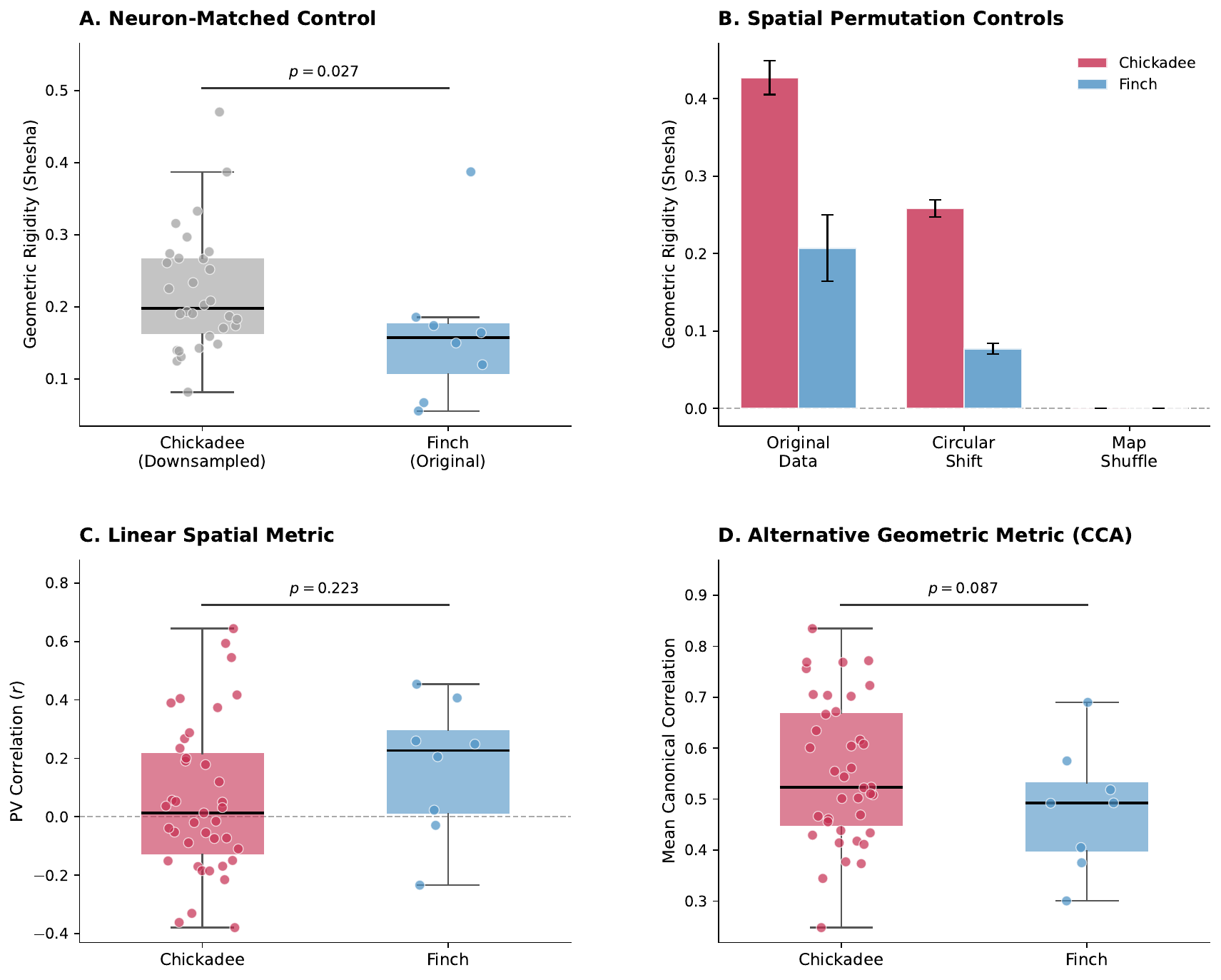}
    \caption{\textbf{Methodological controls and alternative stability benchmarks.}
    \textbf{(A)} Neuron-matched downsampling control. Chickadee ensembles were randomly downsampled to match the median finch neuron count ($N = 6$; 30/39 sessions eligible). The geometric stability advantage persists (Cohen's $d = 0.560$, $p = 0.025$), confirming the effect reflects genuine population structure rather than recording depth.
    \textbf{(B)} Spatial permutation controls. Circular shifting, which preserves individual spatial tuning statistics but misaligns the population vector, substantially reduces geometric stability in both species (chickadee: $0.241 \rightarrow 0.154$; finch: $0.171 \rightarrow 0.072$). Full map shuffling completely abolishes it ($\textrm{Shesha}_{\textrm{FS}} \approx 0.000$), verifying that rigidity requires authentic, coordinated spatial continuity across the full population.
    \textbf{(C)} Split-half population vector (PV) correlation. This traditional linear rate-map metric fails to detect the caching advantage ($p = 0.223$), demonstrating that geometric rigidity is not carried by standard firing rate reliability.
    \textbf{(D)} Canonical Correlation Analysis (CCA) stability. This independent geometry-sensitive measure, quantifying shared subspace structure across neuron splits, shows a trend toward a caching advantage (chickadee $0.554$ vs.\ finch $0.481$, $p = 0.087$), corroborating the Shesha framework and confirming that capacity is encoded in higher-order manifold geometry rather than individual firing rates.}
    \label{fig:supp_controls}
\end{figure}

\newpage

\begin{figure}[H]
    \centering
    \includegraphics[width=\linewidth]{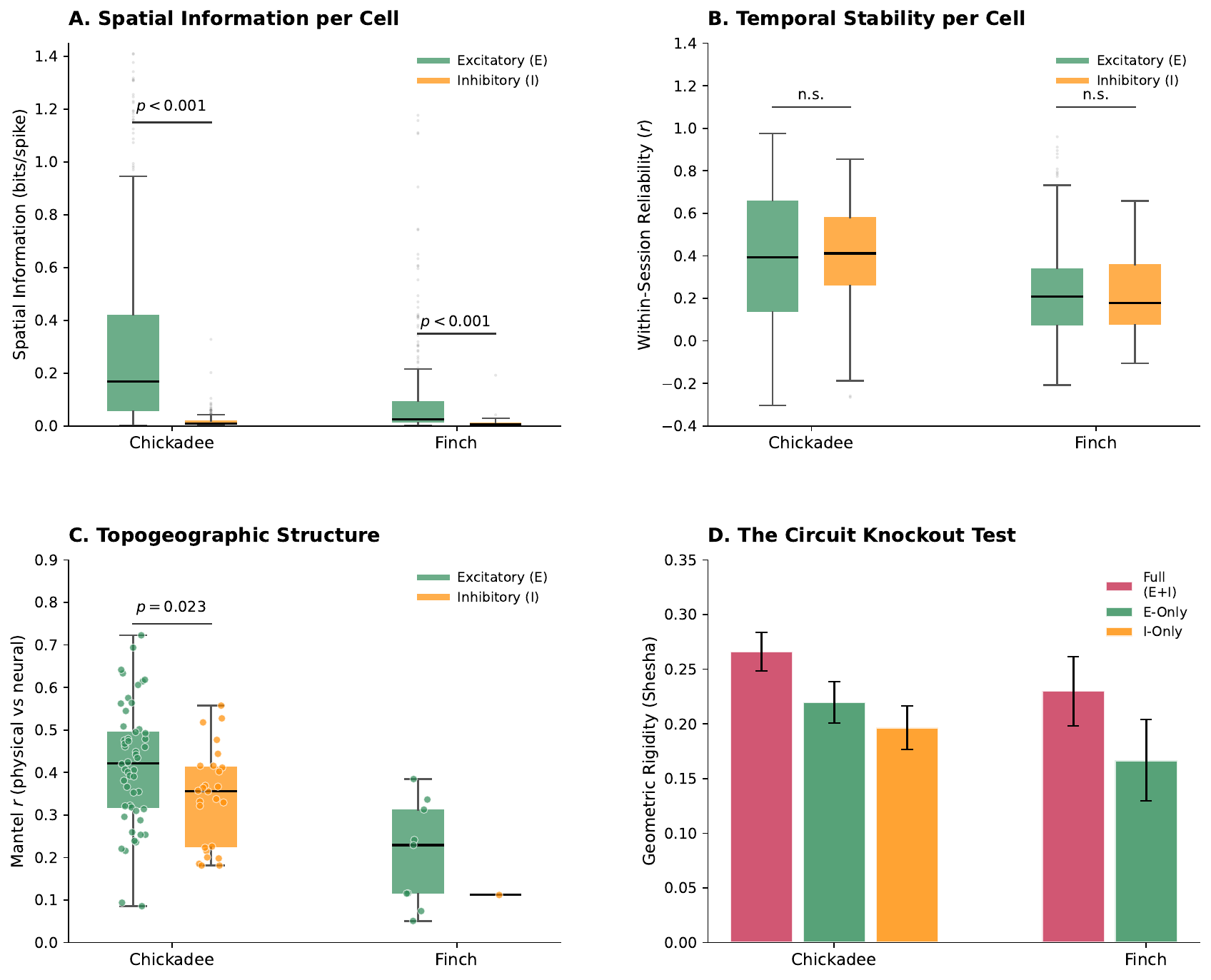}
    \caption{\textbf{Excitatory-inhibitory dissociation of spatial information and temporal stability.} 
    \textbf{(A)} Spatial information content (bits/spike) of individual excitatory (E) and inhibitory (I) neurons. Excitatory cells carry significantly more spatial information than inhibitory cells in both species, confirming E cells as the primary locus of sharp spatial tuning. 
    \textbf{(B)} Within-session temporal stability (split-half correlation of spatial rate maps). Despite lacking sharp spatial tuning, inhibitory cells exhibit temporal reliability that is statistically indistinguishable from excitatory cells, consistent with a robust, broad contextual or gain signal rather than mere recording noise.
    \textbf{(C)} Topogeographic structure evaluated via the Mantel test ($r$). The strict physical-neural structural correspondence is predominantly driven by the excitatory subpopulation. 
    \textbf{(D)} The \textit{in silico} Knockout Test. The absolute geometric rigidity (Shesha) of the network drops significantly when inhibitory neurons are artificially removed from the population code (E-only), demonstrating that intact I-cell dynamics are structurally required to stabilise the high-capacity E-cell geometry.}
    \label{fig:supp_ei_cells}
\end{figure}

\newpage

\begin{figure}[H]
\centering
\includegraphics[width=\textwidth]{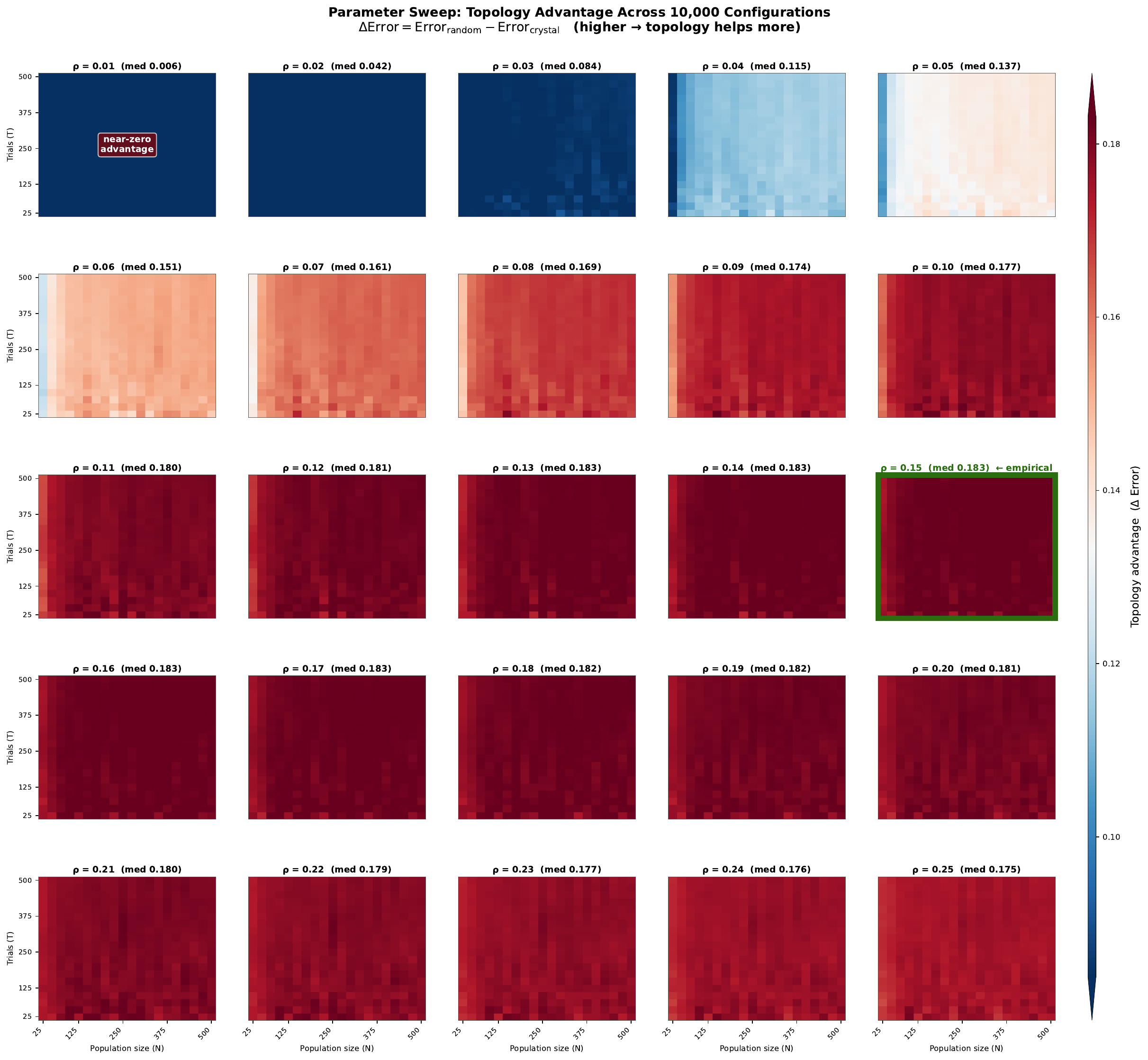}
\caption{\textbf{Topology advantage is robust across 10,000 network
configurations.}
Each panel shows a distinct sparsity level ($\rho = 0.01$--$0.25$); within each panel, the $x$-axis is population size ($N = 25$--$500$ neurons) and the $y$-axis is the number of trials ($T = 25$--$500$). Color encodes the topology advantage $\Delta\mathrm{Error} = \mathrm{Error}_{\mathrm{random}} -
\mathrm{Error}_{\mathrm{crystal}}$, where higher values indicate greater benefit of topological organization over an unstructured network. At very low sparsity ($\rho = 0.01$) the advantage is near zero regardless of $N$ or $T$, indicating that insufficient spatial coverage prevents any code from organizing a stable manifold. The advantage rises steeply between $\rho = 0.02$ and $\rho = 0.10$ and saturates at approximately $\rho = 0.11$ (median $\Delta\mathrm{Error} \approx 0.180$), beyond which further increases in sparsity yield diminishing returns. Critically, the advantage is uniform across the full range of population sizes and trial counts within each panel, demonstrating that the topology benefit is not an artifact of a particular sample-size regime. The highlighted panel (\textbf{green border}, $\rho = 0.15$) corresponds to the empirical sparsity of chickadee hippocampal neurons in the Payne dataset, which falls squarely within the saturated high-capacity regime (median $\Delta\mathrm{Error} = 0.183$). These results confirm that the $>$100-fold capacity advantage reported
in the main text is robust to variation in network scale and data quantity across the full parameter space explored.}
\label{fig:sweep}
\end{figure}

\newpage

\begin{figure}[H]
    \centering
    \includegraphics[width=\linewidth]{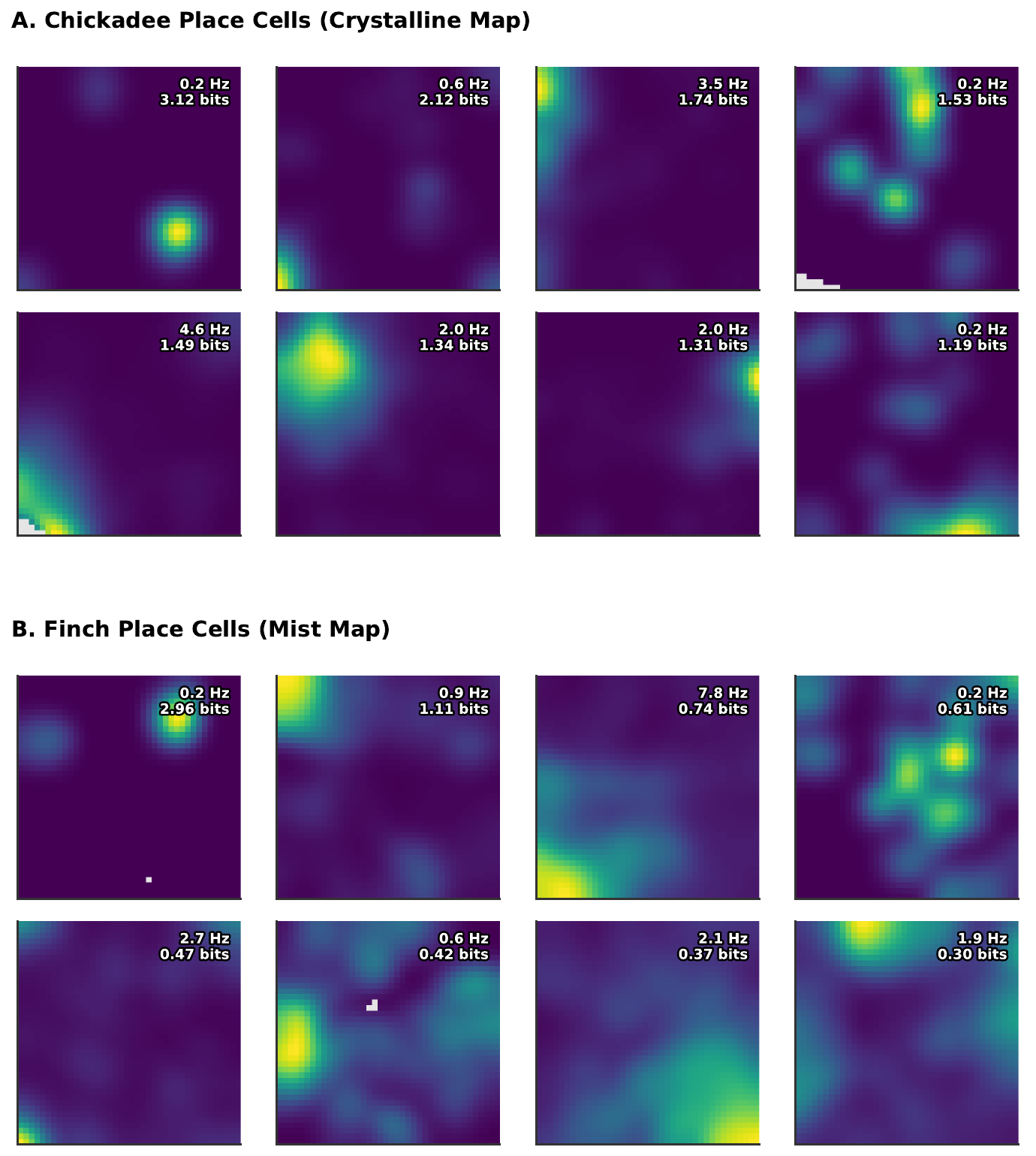}
    \caption{\textbf{Representative raw spatial firing rate maps.} 
    \textbf{(A)} Eight highly informative excitatory neurons from the food-caching chickadee. The rate maps demonstrate the highly heterogeneous, multi-field ``crystalline'' tuning structure characteristic of the caching hippocampus. Unvisited spatial bins are masked in light grey. Peak firing rate (Hz) and spatial information (bits/spike) are inset for each cell.
    \textbf{(B)} Eight highly informative excitatory neurons from the non-caching finch. Even among the most spatially selective cells in the finch population, spatial tuning is characterised by diffuse, single-field or un-tuned ``mist-like'' firing that lacks the sharp topographic boundaries required for a high-capacity rigid manifold.}
    \label{fig:supp_raw_maps}
\end{figure}

\newpage

\begin{figure}[H]
    \centering
    \includegraphics[width=\textwidth]{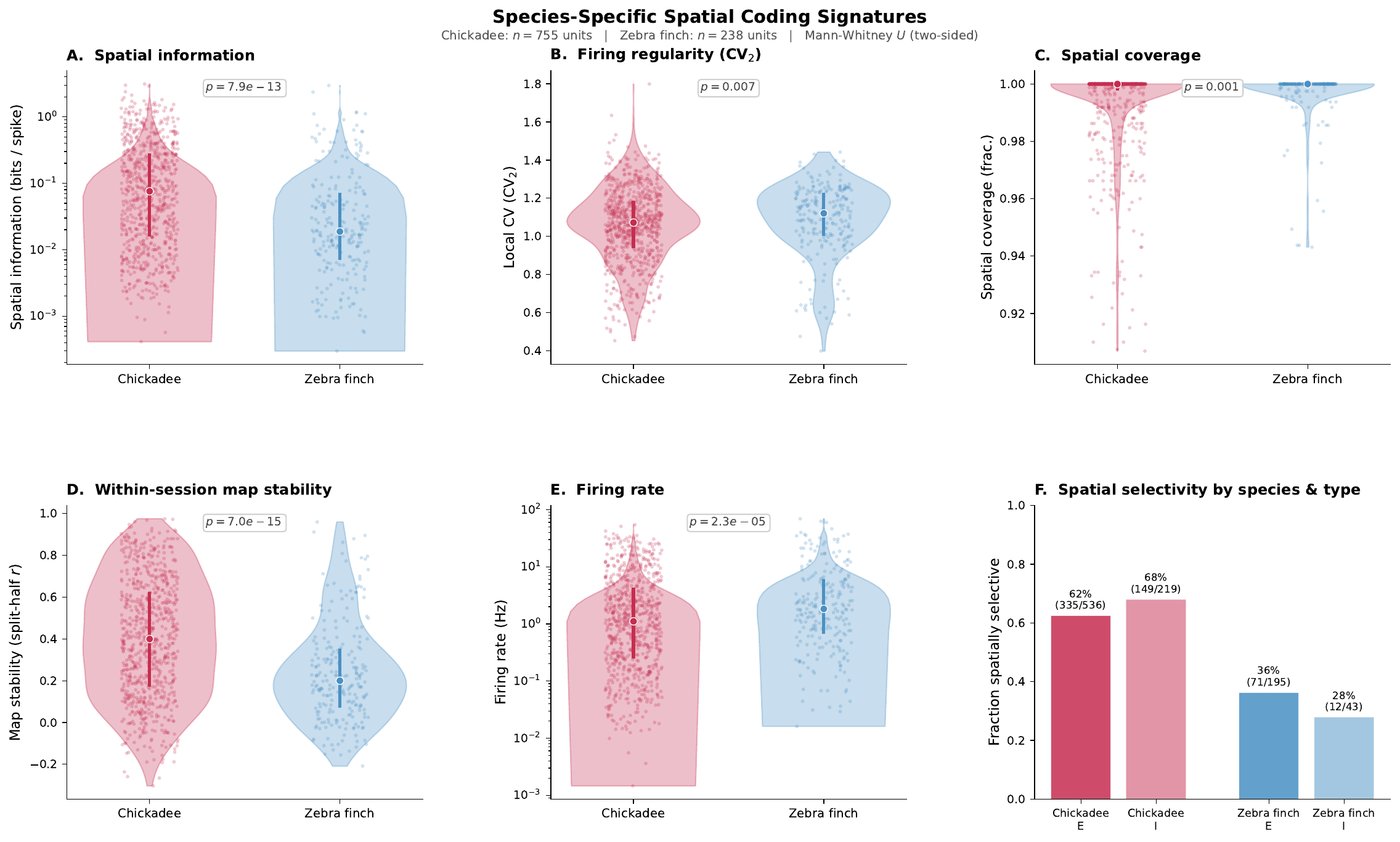}
    \caption{\textbf{Single-unit correlates of population-level geometric stability in food-caching versus non-caching birds.} 
    \textit{In vivo} hippocampal recordings from the food-caching black-capped chickadee (\textit{titmouse}, red; $n = 755$ units) and the non-caching zebra finch (blue; $n = 238$ units) reveal single-unit signatures consistent with the population-level crystalline code identified by Shesha analysis (Figure~\ref{fig:rdm_main}). \textbf{(A)} Spatial information (bits/spike) is substantially elevated in chickadees, consistent with higher per-neuron coding fidelity in the high-capacity regime.
    \textbf{(B)} Firing regularity ($CV_2$) is modestly but significantly elevated in chickadees ($p = 0.007$), reflecting increased temporal precision of place-selective spiking.
    \textbf{(C)} Spatial coverage is near-ceiling for both species, reflecting the open-field recording design rather than a species-specific biological difference.
    \textbf{(D)} Within-session map stability, quantified via split-half spatial map correlation, is markedly higher in chickadees ($p = 7.0 \times 10^{-15}$)---a single-unit analogue of the Shesha effect, confirming that geometric stability is manifest at the level of individual place fields.
    \textbf{(E)} Mean firing rate distributions; chickadees exhibit higher rates on average, consistent with excitatory-dominant drive identified in the circuit analysis (Figure~\ref{fig:ei_scaffold}).
    \textbf{(F)} Fraction of spatially selective units by species and putative cell type. The excitatory--inhibitory selectivity asymmetry is larger in chickadees (E: 62\%, I: 68\%) than in zebra finches (E: 36\%, I: 28\%), consistent with excitatory neurons serving as the primary structural nodes of the crystalline code.
    For violin plots \textbf{(A--E)}, internal dots represent the median, thick vertical lines span the interquartile range, and shaded regions denote kernel density estimates. All comparisons are two-sided Mann-Whitney $U$ tests.}
    \label{fig:empirical_capacity}
\end{figure}

\end{document}